\documentclass[prd,preprintnumbers,superscriptaddress,nofootinbib,floatfix,twocolumn,notitlepage]{revtex4}
\pdfoutput=1
\usepackage{amssymb,amsfonts,amsmath}
\usepackage{graphicx}
\usepackage{ngerman}
\usepackage{color,braket,array,multirow}
\definecolor{darkgreen}{rgb}{0,0.5,0}

\newcommand{\be}{\begin{eqnarray}}
\newcommand{\ee}{\end{eqnarray}}
\def\lsim{\mathrel{\rlap{\lower4pt\hbox{\hskip 0.5 pt$\sim$}}
\raise1pt\hbox{$<$}}}

\newcommand{\cO}{\mathcal{O}}
\newcommand{\cR}{\mathcal{R}}

\begin{document}

\title{Pinning down electroweak dipole operators of the top quark}

\author{Markus Schulze}
\email{markus.schulze@cern.ch}
\affiliation{CERN Theory Division, 1211 Geneva 23, Switzerland.}

\author{Yotam Soreq}
\email{soreqy@mit.edu}
\affiliation{Center for Theoretical Physics, Massachusetts Institute of Technology, Cambridge, MA 02139, U.S.A.}

\begin{abstract}
We consider hadronic top quark pair production and pair production in association with a photon or a $Z$ boson 
to probe electroweak dipole couplings in $t\bar{b}W$, $ t \bar{t}\gamma$ and $ t \bar{t}Z$ interactions.
We demonstrate how measurements of these processes  at the 13\,TeV LHC can be combined to disentangle and constrain anomalous dipole operators.
The construction of cross section ratios allows us to significantly reduce various uncertainties and exploit 
orthogonal sensitivity between the $ t \bar{t}\gamma$ and $ t \bar{t}Z$ couplings.
In addition, we show that angular correlations in $t\bar{t}$ production can be used to constrain the remaining $t\bar{b}W$ dipole operator. 
Our approach yields excellent sensitivity to the anomalous couplings and can be a further step towards precise and direct measurements 
of the top quark electroweak interactions.
\end{abstract}

\preprint{MIT-CTP/4790}
\preprint{CERN-TH-2016-070}
\maketitle

\section{Introduction} \label{sec:intro}

The dynamics of top quark production and decay have been extensively studied at hadron colliders.
Early measurements at the Tevatron have taught us about the  production mechanism of heavy quarks pairs in Quantum Chromodynamics~(QCD)
and later evolved into precision measurements of the top quark mass, spin correlations and the forward-backward asymmetry~\cite{Wicke:2010cg,Deliot:2010ey}.
Similar measurements have been performed during Run-I of the Large Hadron Collider~(LHC) which superseded earlier results at an impressive pace~\cite{Schilling:2012dx,delDuca:2015gca}.
A myriad of measurements has profoundly shaped our understanding of top physics in the Standard Model~(SM) and it led to strong exclusion bounds on new physics. 

Despite this progress, our understanding of the electroweak interactions of the top quark 
({\it i.e.} its coupling to electroweak gauge bosons and the Higgs boson) is rather limited. 
The main reasons are the high production thresholds of the $t\bar{t}+X$ processes, small branching fractions and large backgrounds. 
First measurements and searches at 7\,TeV and 8\,TeV LHC have established some of 
these SM processes~\cite{CMS:2014wma,Khachatryan:2014qaa,Aad:2015iha,Aad:2015gra,Aad:2014lma,Aad:2015uwa,Khachatryan:2015ila,Khachatryan:2015sha,Aad:2015eua}, 
but detailed studies of electroweak couplings will only be possible with a larger data set at higher energies during Run-II. 
These studies will certainly improve our understanding of top interactions with the electroweak sector and potentially probe physics beyond the SM.

The flavor-changing $t\bar{b}W$ interaction is the best-known top quark coupling as it is experimentally accessible through top quark decays 
in $t\bar{t}$ production and single-top quark processes.
This is reflected by precise measurements of $W$ helicity fractions and top quark spin correlations~\cite{Aaltonen:2012rz,Khachatryan:2014vma,Aad:2015bfa,Aad:2015yem}
which can be translated into bounds on anomalous couplings. 
All other electroweak interactions of the top quark with the $Z$, $\gamma$ and the Higgs boson are much less explored.
For example, the top quark electric charge, which governs the coupling strength of the vector-like $t\bar{t}\gamma$ interaction, is known to be 
$Q_t\!=\!+2/3$ with a confidence level larger than $5\,\sigma$~\cite{ATLAS:2011dha,CMS:2012oua,Aaltonen:2013sgl,Abazov:2014lha}. 
However these determinations  were obtained from measuring the electric charges of $W$ boson and $b$-jet in $t\bar{t}$ production, inferring $Q_t\!=\!Q_W+Q_b$.
A hard photon was never present in the event sample and the fundamental $t\bar{t}\gamma$ interaction was not probed.
Similarly, current LHC data only allows constraints on the vector and axial parts of the $t\bar{t}Z$ vertex 
with $\mathcal{O}(100\%)$ uncertainties~\cite{Khachatryan:2015sha}, 
while the respective dipole couplings are unconstrained from hadron collider experiments.  
We note that low-energy observables, such as rare $K$ and $B$ decays~\cite{Grzadkowski:2008mf,Kamenik:2011dk,Brod:2014hsa}, 
together with electroweak precision data \cite{ALEPH:2005ab,Abdallah:2008ab,Larios:1999au,deBlas:2015aea}
can provide strong constraints on modified $t\bar{t}Z$ interactions. 
However, these fairly indirect probes are based on either $Z \to b\bar{b}$ decays or highly off-shell top quarks and $Z$ bosons
in $b \to s \, Z^* \! /\gamma^*$ transitions which rely on additional assumptions on the new physics and are prone to hadronic uncertainties.

This immediately motivates precise studies of the final states $t\bar{t} + Z/\gamma/h$, which yield {\it direct} sensitivity to the desired couplings. 
The commencing Run-II of the LHC with a collision energy of 13~TeV will, for the first time, produce sufficiently many 
events to enable coupling studies.
The two final states considered in this paper, $t\bar{t}+\gamma$ and $t\bar{t}+Z(\to \ell\ell)$, with semi-hadronically decaying top quark pairs,
have typical fiducial cross sections of $\sim400$\,fb and $5$\,fb, respectively.
Hence, we expect about 40,000~$t\bar{t}\gamma$ and 500~$t\bar{t}Z$ events from 100\,fb$^{-1}$ of integrated luminosity. 

\begin{table*}[ht]
\centering
\begin{tabular}{ | c | c | c | c | }
  \multicolumn{1}{l|}{}    &\hspace{1mm}  \includegraphics[width=15mm]{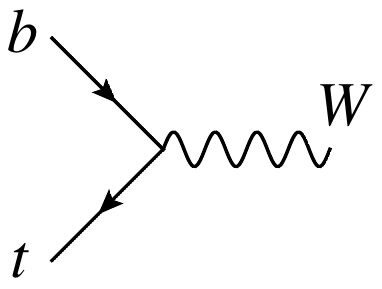} \hspace{1mm}  &  \hspace{1mm}\includegraphics[width=15mm]{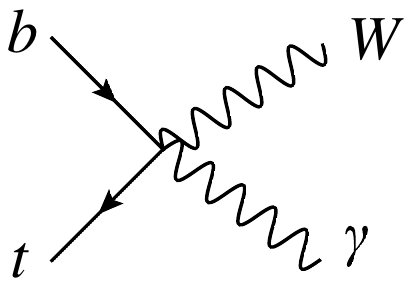}  \hspace{1mm} &  \hspace{1mm}\includegraphics[width=15mm]{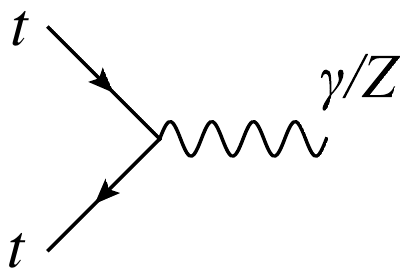} \hspace{1mm}  \\ \hline
\hspace{2mm} $C_{uW}^{33}$   \hspace{2mm}    & $\bigotimes$   & $\bigotimes$  & $\bigotimes$  \\
 $C_{dW}^{33}$       & $\bigotimes$   & $\bigotimes$  &               \\
 $C_{uB\phi}^{33}$   &                &               &  $\bigotimes$ \\
\end{tabular}
\caption{ \label{tab:diags}
Contribution of dimension-six Wilson coefficients to various vertices appearing in $t\bar{t}$, $t\bar{t}+\gamma$ and $t\bar{t}+Z$ production.}
\end{table*}

Anomalous electroweak top quark coupling studies using these processes have been presented in 
Refs.~\cite{Baur:2004uw,Berger:2009hi,AguilarSaavedra:2010nx,AguilarSaavedra:2011ct,Rindani:2011pk,Bach:2012fb,Zhang:2012cd,Aguilar-Saavedra:2014eqa,Zhang:2014rja,Durieux:2014xla,Prasath:2014mfa,Aguilar-Saavedra:2014vta,Bernardo:2014vha,Rontsch:2014cca,Tonero:2014jea,Rontsch:2015una,Buckley:2015lku,Zhang:2016omx,Bylund:2016phk,Cao:2015doa,Cirigliano:2016njn}. 
For example, the TopFitter collaboration \cite{Buckley:2015lku} most recently combined various top quark measurements from the Tevatron and the LHC to 
present a global fit. 
In Ref.~\cite{Aguilar-Saavedra:2014eqa} the authors study dedicated differential observables to probe CP violating top quark decays. 
Anomalous coupling studies that go beyond the leading-order have been presented in Refs.~\cite{Zhang:2014rja,Zhang:2016omx,Rontsch:2014cca,Rontsch:2015una,Bylund:2016phk}.
In this work, we go beyond previous studies by combining observables from $t\bar{t}$, $t\bar{t}+\gamma$ and $t\bar{t}+Z$
to investigate sensitivity to the three electroweak dipole operators which enter simultaneously in these processes. 
We construct cross section ratios to cancel correlated uncertainties and investigate angular asymmetries of the top quark decay products
to boost sensitivity to possible effects of new physics.

\section{Setup} \label{sec:setup}

In the SM, the fundamental interaction vertices of electroweak vector bosons with fermions are given by 
\begin{equation}
  \Gamma_{q'qV}^{ \mathrm{SM}} =  \bar{q}' \; \gamma^\mu \left( d_\mathrm{L}^V P_\mathrm{L} + d_\mathrm{R}^V P_\mathrm{R} \right) q
  \; \varepsilon_\mu^V,
\label{eq:dipolcoupl}
\end{equation}
where $V=\gamma,Z,W^\pm$ and $P_\mathrm{L,R}$ are the left and right-handed chirality projectors. 
The respective couplings $d_\mathrm{L,R}^V$ are fixed by the quantum numbers and gauge symmetries of the SM, see for example Ref.~\cite{AguilarSaavedra:2008zc}.
In this work, we focus on additional contributions from anomalous electroweak dipole moments in the top quark sector.
Their coupling structure is given by 
\begin{equation}
  \delta \Gamma_{q'qV} =  \bar{q}' \; \frac{\mathrm{i} \sigma^{\mu\nu} k_\nu}{m_t} \left( g_\mathrm{L}^V P_\mathrm{L} + g_\mathrm{R}^V P_\mathrm{R} \right) q
  \; \varepsilon_\mu^V,
\label{eq:dipolcoupl}
\end{equation}
where $\sigma^{\mu\nu}=\mathrm{i}/2 \, [\gamma^\mu,\gamma^\nu]$, $k=p_q-p_{q'}$ and $g_\mathrm{L,R}^V$ are the dipole couplings.
The SM has no such interactions at tree level, but electroweak loop corrections radiatively generate dipole moments. 
Their size is well below 1~per-mille~\cite{Bernabeu:1995gs,Czarnecki:1996rx,Hollik:1998vz} which makes them inaccessible by LHC experiments. 
Hence, any sizable deviation from zero will indicate an anomalous interaction from physics beyond the SM.
In order to understand how deep the new physics scales can be probed, we investigate the effects of Eq.~\eqref{eq:dipolcoupl} on physical observables.

Various well-motivated models of new physics~\cite{Hollik:1998vz,Agashe:2006wa,Kagan:2009bn,Ibrahim:2010hv,Ibrahim:2011im,Grojean:2013qca,Richard:2013pwa} 
predict sizable top quark dipole moments.
In this work, we adopt a model-independent approach and assume that the new physics is CP conserving and respects the full SM gauge symmetry.
We use the effective field theory parameterization of Ref.~\cite{AguilarSaavedra:2008zc} in terms of higher dimensional 
operators $\mathcal{L}^\mathrm{dim6} = \sum_i C_i /\Lambda^2 \, \mathcal{O}_i $ and a new physics scale $\Lambda$. 
The relevant operators in our analysis are 
\begin{align}   
        \label{eq:OuW}
        \cO^{33}_{uW} &= \left( \bar{q}_{\mathrm{L}} \sigma^{\mu\nu}\tau^I t_{\mathrm{R}} \right) \tilde{H} W^I_{\mu\nu} , \\
        \label{eq:OdW}
        \cO^{33}_{dW} &= \left( \bar{q}_{\mathrm{L}} \sigma^{\mu\nu}\tau^I b_{\mathrm{R}} \right) H W^I_{\mu\nu} , \\
        \label{eq:OuB}
        \cO^{33}_{uB\phi} &=  \left( \bar{q}_{\mathrm{L}} \sigma^{\mu\nu} t_{\mathrm{R}} \right) \tilde{H} B_{\mu\nu},
\end{align}
where $\tilde{H} = \mathrm{i} \, \tau^2H^*$, $H=(0,h+v)/\sqrt{2}$ is the Higgs boson doublet with 
the vacuum expectation value $v=246\,$GeV. $B_{\mu\nu}(W^I_{\mu\nu})$ is the 
$\mathrm{U}(1)_\mathrm{Y}(\mathrm{SU}(2)_\mathrm{L})$ gauge 
field signal strength and $I$ is adjoint $\mathrm{SU}(2)_\mathrm{L}$ index. 
$q_\mathrm{L}=(t_\mathrm{L},b_\mathrm{L})$, $b_\mathrm{R}(t_\mathrm{R})$ are the quark left-handed doublet 
and bottom\,(top) right-handed singlet, respectively.  
Using the parameterization of Eq.~\eqref{eq:dipolcoupl} and the operators in Eqs.~\eqref{eq:OuW}--\eqref{eq:OuB} we find 
\begin{align}
\label{eq:EFTcoupl}
  g_\mathrm{L}^{W^-} &=g_\mathrm{R}^{W^+ *} 
 		 = -\frac{e \, m_t}{s_\mathrm{W} M_W } \frac{v^2}{\Lambda^2} \, C^{33*}_{dW} ,  \\
  g_\mathrm{R}^{W^-} &= g_\mathrm{L}^{W^+ *}
  		= -\frac{e \, m_t}{s_\mathrm{W} M_W } \frac{v^2}{\Lambda^2} \, C^{33}_{uW} ,   \nonumber \\
  g_\mathrm{L}^{\gamma} &= g_\mathrm{R}^{\gamma *} 
                        = -\frac{\sqrt{2} \, m_t \, v}{\Lambda^2} \left( c_\mathrm{W} C^{33*}_{uB\phi} + s_\mathrm{W} C^{33*}_{uW} \right),
  \nonumber \\
  g_\mathrm{L}^{Z} &= g_\mathrm{R}^{Z *} 
                   = -\frac{e  \, m_t \, v^2}{\sqrt{2} s_\mathrm{W} c_\mathrm{W} M_Z \Lambda^2} \left( c_\mathrm{W} C^{33*}_{uW} - s_\mathrm{W} C^{33*}_{uB\phi}  \right),
  \nonumber
\end{align}
where $e$ is the electric coupling and $s_\mathrm{W}\,(c_\mathrm{W})$ is sine\,(cosine) of the weak mixing angle.
We emphasize that since the various dipole couplings in $t\bar{b}W$, $t \bar{t} \gamma/Z$ are related via the underlying $SU(2)\times U(1)$ gauge invariance, the coupling degrees of freedom are reduced to only three Wilson coefficients $C^{33}_{uW}$, $C^{33}_{dW}$ and $C^{33}_{uB\phi}$.
In the expression for $g_{\mathrm{L,R}}^{\gamma / Z}$, we see the characteristic weak mixing angle rotation 
between $C^{33}_{uW}$ and $C^{33}_{uB\phi}$ which will be responsible for the {\it orthogonal} constraints 
that we find in $t\bar{t}+\gamma$ and $t\bar{t}+Z$ production below. 

At this point we note that there is one more allowed Lorentz-structure in addition to Eq.~(\ref{eq:dipolcoupl}), 
which is relevant for our analysis,
\begin{align}
 \delta \Gamma_{q' q V V'} =  \bar{q}' \; \mathrm{i} \sigma^{\mu\nu} \left( g_\mathrm{L}^{V V'} P_\mathrm{L} + g_\mathrm{R}^{V V'} P_\mathrm{R} \right) q
  \; \varepsilon_\mu^V \; \varepsilon_\nu^{V'}.
\label{eq:dipolcoupl2}
\end{align}
This four-point vertex enters our calculation only in one specific place: the radiative top quark decay $ t \to bW +\gamma $ of the $t\bar{t} \gamma$ final state.
In a complete description of $t\bar{t} \gamma$ production, the photon can arise from either the hard production stage ({\it before the top quarks go on-shell}), 
or the top quark decay stage ({\it after one top quark went on-shell}).
The latter part constitutes more than $50\%$ of the total $t\bar{t} \gamma$ cross section~\cite{Melnikov:2011ta} 
and receives contributions from Eq.~(\ref{eq:dipolcoupl2}).
It is therefore crucial to account for this contribution in the analysis of $t\bar{t} \gamma$ final states\footnote{
The effects from radiative top quark decays are irrelevant for the $t\bar{t}+Z$ process because the 
decay $t \to bW +Z$ is suppressed by a phase space factor of $10^{-6}$.}.
The two additional dipole couplings in Eq.~(\ref{eq:dipolcoupl2}) are given by
\begin{align}
\label{eq:EFTcoupl2}
   g_\mathrm{L,R}^{W^\sigma \gamma} = (-e) \,  \sigma \, g_\mathrm{L,R}^{W^\sigma} ,
\end{align}
where $\sigma=\pm1$ is the electric charge of the respective $W$ boson.
In Table~\ref{tab:diags} we summarize the anomalous coefficients considered in our analysis and their appearance in various interaction vertices.

The up-to-date hadron collider bounds on the Wilson coefficients of Eqs.~(\ref{eq:OuW}--\ref{eq:OuB}) are summarized in Ref.~\cite{Buckley:2015lku} and found to be
\begin{align}
        C^{33}_{uW} &\in [-4.0,\,3.4] \, (\Lambda/{\rm TeV})^2, \\
        C^{33}_{uB\phi} &\in [-7.1,\, 4.7] \, (\Lambda/{\rm TeV})^2
\end{align}
at 95\,\% confidence level~(CL). 
Note that in order to infer the bound on $ C^{33}_{uW}$, 
the contributions of all other top-related dimension six operators were marginalized, 
while for $C^{33}_{uB\phi}$, it was assumed that the only new physics contribution is from $\cO^{33}_{uB\phi}$. 
Finally, in Ref.~\cite{AguilarSaavedra:2011ct} we find that
\begin{align}
        C^{33}_{dW} \in [-2.3,\,+2.3] \, (\Lambda/{\rm TeV})^2, \quad
\end{align} 
at 95\,\%~CL. 
Note that flavor violating contributions from $\cO^{33}_{uB\phi}$ can be avoided by alignment to the up sector in flavor space. 
In that limit, $\cO^{33}_{uW}$ induces an irreducible source of 
flavor violation in the charged current, which is however suppressed by off-diagonal CKM matrix elements, for a more detailed discussion see Ref.~\cite{Fox:2007in}.
Alignment to the down basis will reduce flavor violation originating from $\cO^{33}_{dW}$, see also the discussion in Ref.~\cite{Drobnak:2011aa}.
\\

In a strict ${1}/{\Lambda^2}$-expansion within an effective field theory even more operators contribute to the processes considered here.
For example, the operator $\bar{\ell} \gamma^\mu \ell \; \bar{t} \gamma_\mu t$ can enter the $t\bar{t}+Z(\to \ell \ell)$
process. 
However, these operators do not exhibit a Breit-Wigner peak structure such as the rest of the amplitude and 
contribute as a fairly constant function of $m_{\ell\ell}$ around the $Z$ boson mass window of $\pm 10$~GeV, see e.g. Fig.~5 in~\cite{Durieux:2014xla}.
Hence, their interference at $\cO (\Lambda^{-2})$ integrates to zero and only their squared contribution at $\cO (\Lambda^{-4})$ survives. 
We therefore neglect all operators with fermionic contact interactions. 
Potential contributions from chromo-magnetic and chromo-electric dipole moments can enter our processes through anomalous top-gluon couplings. 
However, in this work, we assume that QCD is unaltered by new physics and we do not consider chromo-dipole moments. 
If QCD is truly non-fundamental, these anomalous interactions are best searched for in jet processes, 
$gg \to H$ or $t \bar{t}$ production, see e.g. Ref.~\cite{Aguilar-Saavedra:2014iga}.

Besides dipole interactions, also the strength of the $t\bar{t}Z$ and $t\bar{b}W$ vector and axial couplings can be altered by new physics.
In this work we assume their SM value which is a restrictive assumption on new physics scenarios.
Let us therefore comment on possible strategies for a broader analysis in future extensions of our work.
One straight-forward solution is to consider the full space of couplings in a 6-dimensional analysis
(anomalous vector and axial-vector couplings introduce three more operators to the ones considered here  \cite{AguilarSaavedra:2008zc}). 
This approach might however be challenging given the large number of degrees of freedom and the limited number of events. 
Another solution may be a careful analysis of kinematics in sequential steps: 
(i)~A first analysis of the process $pp\to t\bar{t}+\gamma$ can yield information on the dipole moments. 
Anomalous vector or axial-vector couplings cannot develop thanks to gauge symmetries.
(ii)~Any deviation in $t\bar{t}+\gamma$ immediately predicts anomalies in the $pp\to t\bar{t}+Z$ process, as can be seen from Eq.~\eqref{eq:EFTcoupl}.
In Ref.~\cite{Rontsch:2015una} it has been shown that dipole couplings most prominently manifest in energy related contributions such as $p_{\mathrm{T},Z}$. 
Hence, any remaining discrepancy from anomalous vector and axial-vector couplings can be detected in angular distributions, such as $\Delta \phi_{\ell\ell}$, 
which are sufficiently independent of energies. 
(iii)~Once anomalous dipole couplings in $t\bar{t}+\gamma /Z$ are established, only one remaining dipole operators in $t\bar{b}W$ interactions remains and can be constrained as
outlined in Sect. IV. 
Additional anomalous vector and axial couplings in the $t\bar{b}W$ interaction can then be constrained by relating analyses of 
top quark pair and single-top quark production $q b \to W \to  q' t$ \cite{AguilarSaavedra:2011ct}.
Finally, we note that scenarios of large CP-violating electric dipole moments have been extensively studied in the literature~\cite{Bernreuther:1992dz,HernandezSanchez:2006sw,Agashe:2006wa,Ibrahim:2010hv,Dekens:2014jka}. These effects arise through CP violation in loop contributions and we expect them to similarly affect the magnetic dipole moments.
\\

Our description of the processes $t\bar{t}$, $t\bar{t}+\gamma$ and $t\bar{t}+Z(\to \ell\ell)$ treats top quarks in the narrow-width approximation
and includes the full decay chain of the top quarks into a single lepton plus jets final state.
All spin-correlations are retained. 
In $t\bar{t}+\gamma$ final states, the photon is allowed to be emitted in the top quark production stage as well as in the decay stage (including the $W$ and $W$ decay products).
The Monte Carlo simulation is based on the TOPAZ code \cite{RefTOPAZ} developed in Refs.~\cite{Melnikov:2009dn,Rontsch:2015una,Rontsch:2014cca}, 
which we extended to handle all anomalous couplings needed in this analysis.
If not otherwise stated, we use the following generic selection cuts for the $t\bar{t}$, $t\bar{t}+\gamma$ and $t\bar{t}+Z$ processes
\begin{align}
\label{eq:cuts}
   & p_\perp^\ell \ge 20 \, \text{GeV},         & |y_\ell| \le 2.5,  \quad\quad    & E_\perp^\text{miss} \ge 20\, \text{GeV},
   \nonumber \\
   & p_\perp^{j,b} \ge 20 \, \text{GeV},        & |y_j| \le 2.5,     \quad\quad    & |y_b| \le 2.0.
\end{align}
Jets are defined by the anti-$k_\mathrm{T}$ jet algorithm \cite{Cacciari:2008gp} with $R=0.4$.
In addition, for $t\bar{t}+Z$ production we require an invariant mass cut of $|m_{\ell\ell}-M_Z|\le 10$\,GeV.
For $t\bar{t}+\gamma$ we require the isolation cuts $R_{\gamma j}=R_{\gamma \ell}=0.4$ for photons with $p_\perp^\gamma \ge p_{\perp,\mathrm{cut}}^\gamma=20$~GeV 
and $|y_\gamma| \le 2.5$.
In accordance with the findings at NLO QCD \cite{Melnikov:2011ta,Rontsch:2015una}, we set the central renormalization and factorization 
scales to $\mu=m_t$ for $t\bar{t}$ and $t\bar{t}+\gamma$ production,
and $\mu=m_t+M_Z/2$ for $t\bar{t}+Z$ production.
We use NNPDF3.0 \cite{Ball:2014uwa} parton distribution functions, $m_t=173.2$\,GeV, $M_Z=91.1876$\,GeV, $M_W=80.399$\,GeV, 
$G_\mathrm{F}=1.16639 \times 10^{-5}$\,GeV$^{-2}$, 
and use $\alpha=1/137$ in the $t\bar{t}\gamma$ process.

\section{Cross section ratios} \label{sec:ratios}

\begin{figure*}[ht]
\centering
\includegraphics[scale=0.36]{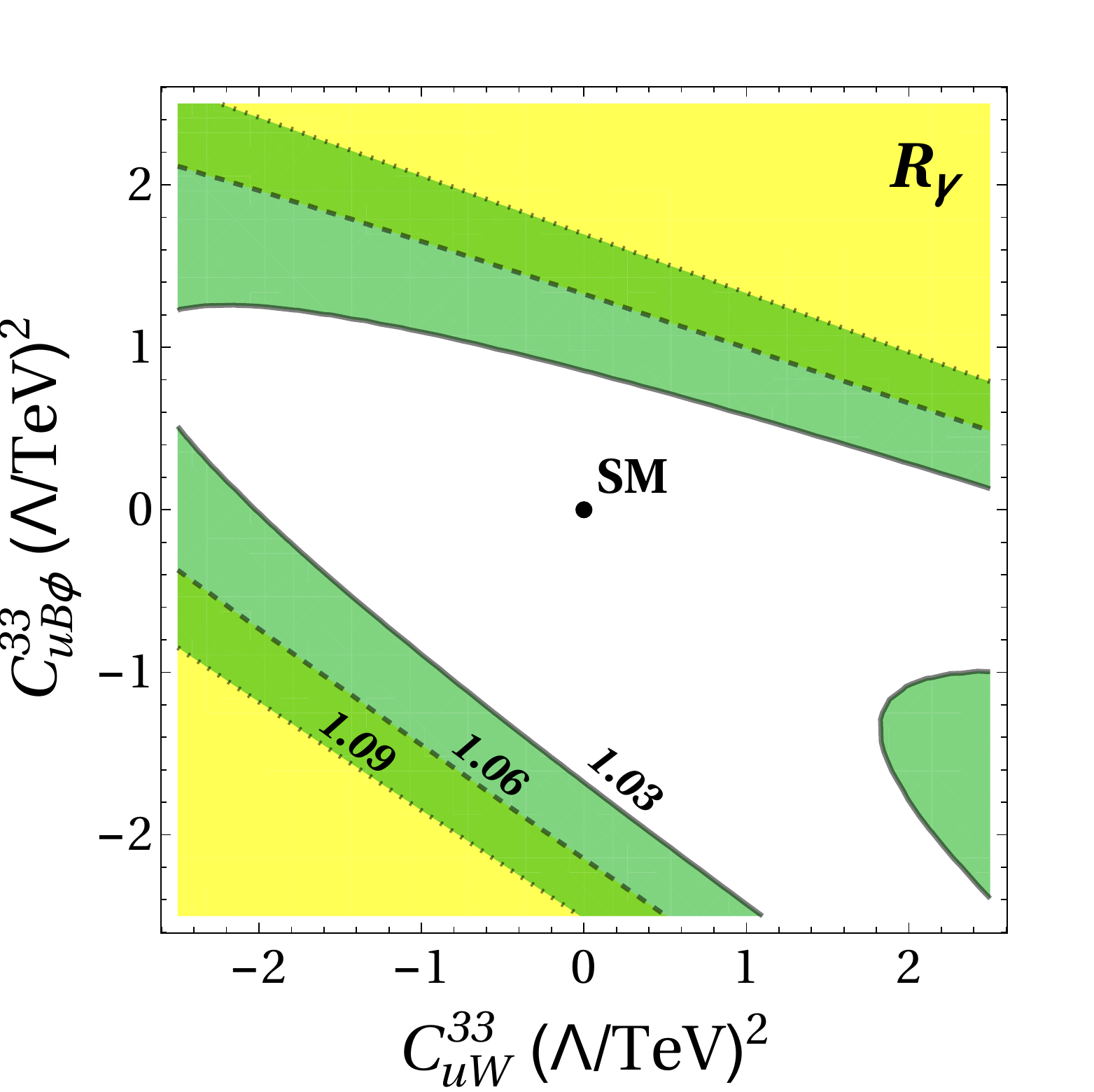}
\hfill
\includegraphics[scale=0.36]{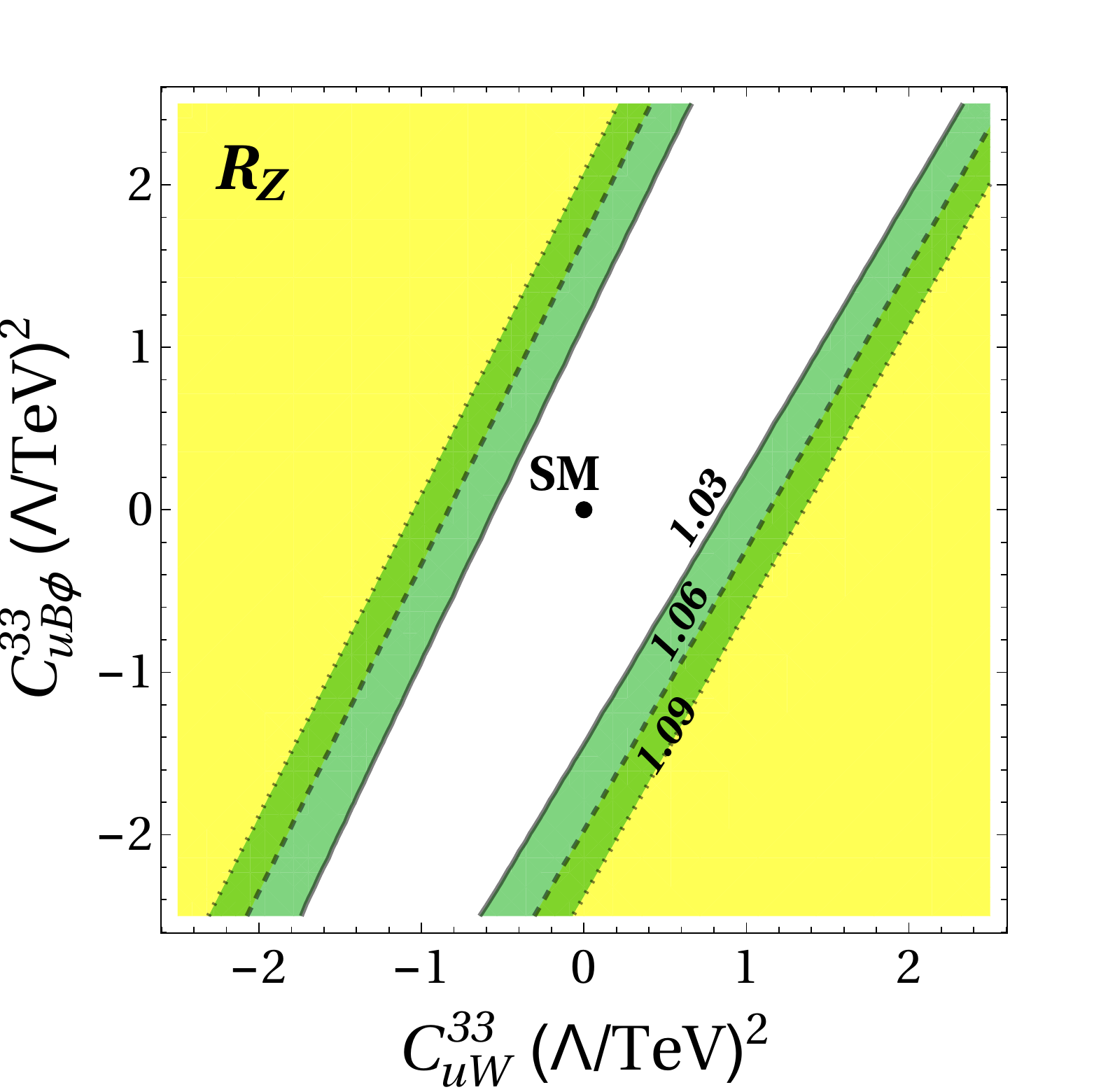}
\hfill
\includegraphics[scale=0.36]{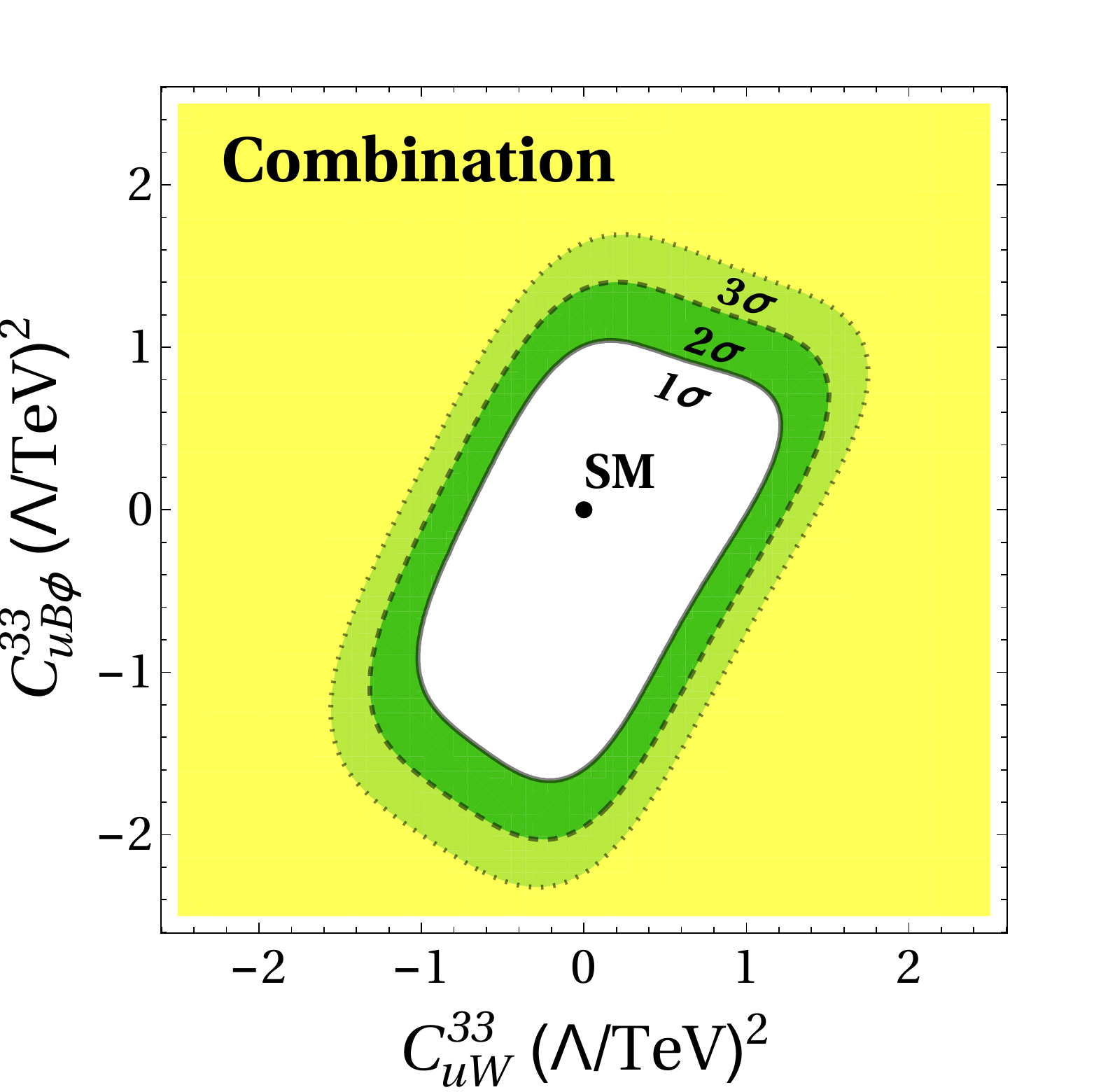}
\caption{ \label{fig:ratioconstraints}
Cross section ratios $\cR_{\gamma}$ (left) and $\cR_{Z}$ (middle) normalized to their SM values ($\cR^{\rm SM}_{\gamma/Z}$) 
as a function of the anomalous dipole operator couplings.
The contours show the deviation from the SM value in steps of 3, 6 and 9 percent.
On the right we show the $1,2,3\sigma$ contours from combining $\cR_\gamma$ and $\cR_{Z}$ 
with an assumed uncertainty of $\Delta\cR_{Z/\gamma}=3\%$.}
\end{figure*}

In the following we study the sensitivity of different cross section ratios to the anomalous dipole couplings. 
Ratios of observables have the advantage that leading uncertainties on e.g. $\alpha_s$ and parton distribution functions~(PDFs) largely cancel.
Even higher order corrections are expected to cancel to some extent, provided that the cross sections are probed in similar regions of phase space.
Experimental uncertainties related to luminosity and jet energy scales drop out in the ratio, as well.

A similar idea of employing ratios has been presented in Ref.~\cite{Plehn:2015cta} for measuring the top quark Yukawa coupling at the 100\,TeV Future Circular Collider.
The authors consider the cross section ratio of $t\bar{t}+H$ over $t\bar{t}+Z$ and demonstrate that a precision of
1\% can be reached. 
This certainly extreme precision is obtained thanks to the kinematic similarities of the two processes and the 
enormous event rate at a 100~TeV collider. 
Hence, in the context of this work it seems suggestive to study the ratio $\sigma_{t\bar{t}Z} / \sigma_{t\bar{t}\gamma}$. 
Yet, we refrain from doing so and instead construct two ratios 
\begin{equation}
   \cR_{\gamma} = \sigma_{t\bar{t}\gamma} \big/ \sigma_{t\bar{t}}, 
   \quad
   \cR_{Z} = \sigma_{t\bar{t}Z} \big/ \sigma_{t\bar{t}},    
\label{eq:ratiodef}
\end{equation}
with respect to the $t\bar{t}$ cross section. This allows to benefit from the large $\sigma_{t\bar{t}}$ cross section which has almost no statistical uncertainties.
Moreover, we will find that $\sigma_{t\bar{t}Z}$ and $\sigma_{t\bar{t}\gamma}$ have orthogonal dependence 
on the anomalous couplings which is distinctly exposed in the ratios of Eq.~(\ref{eq:ratiodef}).
Of course, it is no longer given that uncertainties cancel in these ratios because the $t\bar{t}+\gamma/Z$ and $t\bar{t}$ processes
probe very different energies and phase spaces.
For example, using the cuts in Eq.~(\ref{eq:cuts}) we find at leading order an average center-of-mass energy of 
$\braket{\hat{E}}\approx 525$~GeV for the $t\bar{t}$ process, but $\braket{\hat{E}}\approx 630$~GeV and $860$~GeV for  $t\bar{t}+\gamma$ and  $t\bar{t}+Z$, respectively.
Hence, the parton distribution functions are evaluated at significantly different values of $Q^2$ and a cancellation of uncertainties is not guaranteed.
We circumvent this issue by applying additional (mild) invariant mass cuts on the two $t\bar{t}$ cross sections in Eq.~(\ref{eq:ratiodef})
to increase the average center-of-mass energy. 
In particular, we request $m_{t\bar{t}} \ge 470$~GeV for $\sigma_{t\bar{t}}$ in $\cR_{\gamma}$, and 
$m_{t\bar{t}} \ge 700$~GeV for $\sigma_{t\bar{t}}$ in $\cR_{Z}$.
We verified that the average center-of-mass energy in these two $t\bar{t}$ processes matches the one of the respective $t\bar{t}+\gamma/Z$ process. 
We note that the experimental reconstruction of the $t\bar{t}$ invariant mass is known to involve large systematic errors.
One could therefore worry how this affects the above mentioned cut on $m_{t\bar{t}}$ and how it may spoil the cancellation of uncertainties in the cross section ratios.
We therefore investigated variations of this cut and its impact on the ratios $\cR_{\gamma}$ and $\cR_{Z}$.
While the absolute values obviously change with different values of the cut, we find that our uncertainty estimates presented below are completely unchanged.
Even in the extreme case where we remove the $m_{t\bar{t}}$ cuts entirely, our conclusions remain the same as presented below.

To explicitly quantify cancellations of uncertainties in the ratios, 
we evaluate the cross sections at NLO QCD and study variations of parton distribution functions.
This also allows us to obtain error estimates that will be important when estimating sensitivity to the dipole couplings.
We partially use results from the NLO QCD computations of $t\bar{t}+\gamma$ and $t\bar{t}+Z$ in Refs.~\cite{Melnikov:2011ta,Rontsch:2014cca} 
and recompute the NLO $t\bar{t}$ cross sections
with the same input parameters and the above mentioned cut on $m_{t\bar{t}}$.
The higher average center-of-mass energy of the $t\bar{t}$ cross sections also calls for adapting the renormalization and factorization scales. 
We find it natural to choose $\mu=m_t + p_{\perp,\mathrm{cut}}^\gamma$ and $\mu=m_t + M_Z/2$ for the two $t\bar{t}$ cross sections in Eq.~(\ref{eq:ratiodef}), respectively.
Using this setup, we find
\setlength{\extrarowheight}{0pt}
\begin{align}
     \cR^{\rm SM}_{\gamma} \times 10^{-3} &= 
     \begin{cases}
     11.4^{-0.7\%}_{+0.7\%} \quad \text{at LO},
     \\[1ex]
     12.6^{+3.1\%}_{-1.8\%} \quad \text{at NLO QCD},
     \end{cases}
\\[2ex]
     \cR^{\rm SM}_{Z} \times 10^{-4} &= 
     \begin{cases}
     2.27^{-1.7\%}_{+2.0\%} \quad \text{at LO},
     \\[1ex]
     1.99^{-1.9\%}_{+2.8\%} \quad \text{at NLO QCD}.
     \end{cases}
\label{eq:ratioscalevar}
\end{align}
The upper (lower) values correspond to the lower (upper) scale variation by a factor of two around the respective central scale. 
We observe a scale dependence of $\pm 1\%$ and $\pm 2\%$ at LO for $\cR_{\gamma}$ and $\cR_{Z}$, respectively.
These values are slightly increased to $\pm (2-3)\%$ at NLO QCD. 
This constitutes a remarkable stability with respect to scale variation 
when compared to the cross sections themselves which exhibit a dependence of about $\pm 20\%$ at NLO.
It should be noted that the LO scale variation is far outside the NLO result. 
This is to be expected as $\alpha_\mathrm{s}(\mu_\mathrm{R})$ cancels exactly and the only remaining source of scale
dependence arises from unmatched $q^2$-dependence in the parton distribution functions.
Only our NLO results develop, for the first time, logarithms of the scales with process-dependent coefficients that
yield a good error estimate. 
Hence only the NLO result should be considered a physical meaningful prediction.
For our coupling constraints below we will choose the largest value of the NLO scale variation, $\Delta\cR_{Z/\gamma}=3\%$, as our uncertainty for both ratios.

Given the fiducial cross sections of about $\sim 5\,(400)\,$fb for $t\bar{t}Z\,(t\bar{t}\gamma)$ production, 
the statistical error is expected to be sub-dominant after an integrated luminosity of about $250$\,fb$^{-1}$. 
This argument is supported by a first measurement of $\cR_{\gamma}$ by the CMS collaboration~\cite{CMS:2014wma} 
at 8~TeV. They find the value $\cR_{\gamma}(8\,\mathrm{TeV})=10.7 \times 10^{-3} \pm 6.5\% (\mathrm{stat.}) \pm 25\% (\mathrm{syst.}) $
from an integrated luminosity of 19.7\,fb$^{-1}$.
The dominant systematics arise from background modeling ($\pm 23\%$) which have the potential to be improved in future analyses.

We also investigate uncertainties from parton distribution functions and find the following
results from using three different sets of parton distribution functions:
\begin{align}
     \mathcal{R}_{\gamma}^\mathrm{LO} \times 10^{-3} =& 
     \begin{cases}
     11.5 \quad \text{with NNPDF3.0 \cite{Ball:2014uwa}},
     \\
     11.4 \quad \text{with CTEQ6L1 \cite{Pumplin:2002vw}},
     \\
     11.5 \quad \text{with MSTW08 \cite{Martin:2009iq}},
     \end{cases}
     \\
     \mathcal{R}_{Z}^\mathrm{LO} \times 10^{-4} =& 
     \begin{cases}
     2.29 \quad \text{with NNPDF3.0},
     \\
     2.27 \quad \text{with CTEQ6L1},
     \\
     2.27 \quad \text{with MSTW08}.
     \end{cases}
\end{align}
We observe very stable results with variations at the level of $1\%$, which have to be compared to 
$\pm 10\%$ variations on the cross sections themselves.
Again, we find confirmation that the ratios are true precision observables. 

Finally, let us briefly comment on the impact of electroweak corrections.
Compared to the QCD corrections, they are expected to be much less universal for the $t\bar{t}$ and $t\bar{t}+\gamma/Z$ processes.
Hence, on the one hand, a cancellation is most likely incomplete. 
On the other hand, the electroweak corrections on the cross sections are known to be in the few percent range ($\mathcal{O}(\alpha)$)~\cite{Kuhn:2013zoa,Frixione:2015zaa}.
This leads to a (minor) shift in the absolute values of the ratios but it does not at all affect our estimate of uncertainties. 
\\

Let us now turn to studying the effects of anomalous electroweak dipole moments on the cross section ratios.
Given two (pseudo-)observables, $\cR_{\gamma}$ and $\cR_{Z}$, we investigate their dependence on the Wilson coefficients
$C^{33}_{uW}$ and $C^{33}_{uB\phi}$ (neglecting operator mixing and running from beyond the LO).
The remaining coefficient $C^{33}_{dW}$ we examine through angular asymmetries in $t\bar{t}$ production in the following section.
Since only total cross sections enter this analysis, we are not sensitive to the tail of energy-related distributions where 
possible issues with unitarity violating EFT operators could appear. 
We vary the numerical values of $C^{33}_{uW}$ and $C^{33}_{uB\phi}$ between $[-4,\,4]$ in steps of $1$ and compute 81 LO cross sections 
for each of the $t\bar{t}$, $t\bar{t}+\gamma$ and $t\bar{t}+Z$ processes.
We then perform a two-dimensional analytic fit of the ratios to present our results.
Figure~\ref{fig:ratioconstraints} shows the deviation of the anomalous ratios from the SM value 
for $t\bar{t}\gamma$ on the left and $t\bar{t}Z$ in the middle. 
These two plots strikingly show the {\it orthogonal} dependence of the two ratios on the dipole couplings.
This is a consequence of the already mentioned $s_\mathrm{W}$ rotation pattern of $g^{\gamma,Z}_\mathrm{L,R}$ in Eq.~\eqref{eq:EFTcoupl}.
The white\,(dark green) bands in left and middle plot of Fig.~\ref{fig:ratioconstraints} indicate the parameter spaces 
where the anomalous cross section ratios deviate from the SM by less
than the assumed 1\,(2) standard deviation of the assumed uncertainty. 
Hence, all anomalous couplings outside the dark green bands can be excluded if the measurement is in agreement with the SM at the $2\sigma$ level.
In the third plot, on the right of Fig.~\ref{fig:ratioconstraints}, we show the combination of the two constraints using a naive $\chi^2$ combination.
This noticeably leads to a striking improvement of the constraints as the {\it orthogonal} dependence on the Wilson coefficients
allows us to completely bound the parameter space without a blind direction.
We note that these projected bounds are stronger by factors of 2-3 compared to the current bounds from 7 and 8~TeV cross section measurements. 
Moreover, a better understanding of the anomalous $pp \to b\bar{b} \ell\nu jj + \gamma$ process beyond the LO will further improve these results. 
Such a calculation is currently not available and we refer to future work.

\section{Angular asymmetries in $t\bar{t}$ production} \label{sec:angles}

\begin{figure*}[ht]
 \includegraphics[scale=0.54]{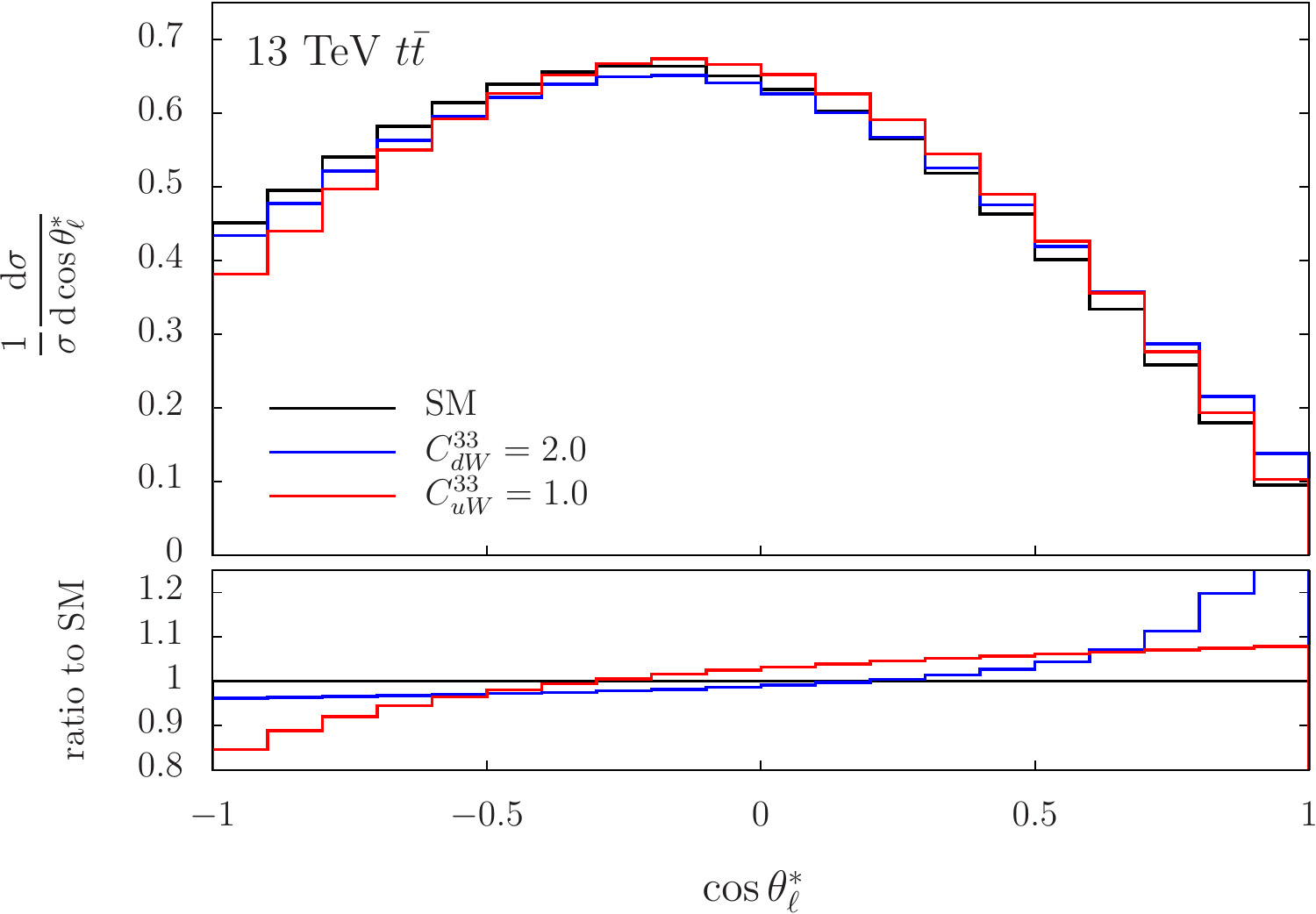} \hfill
 \includegraphics[scale=0.54]{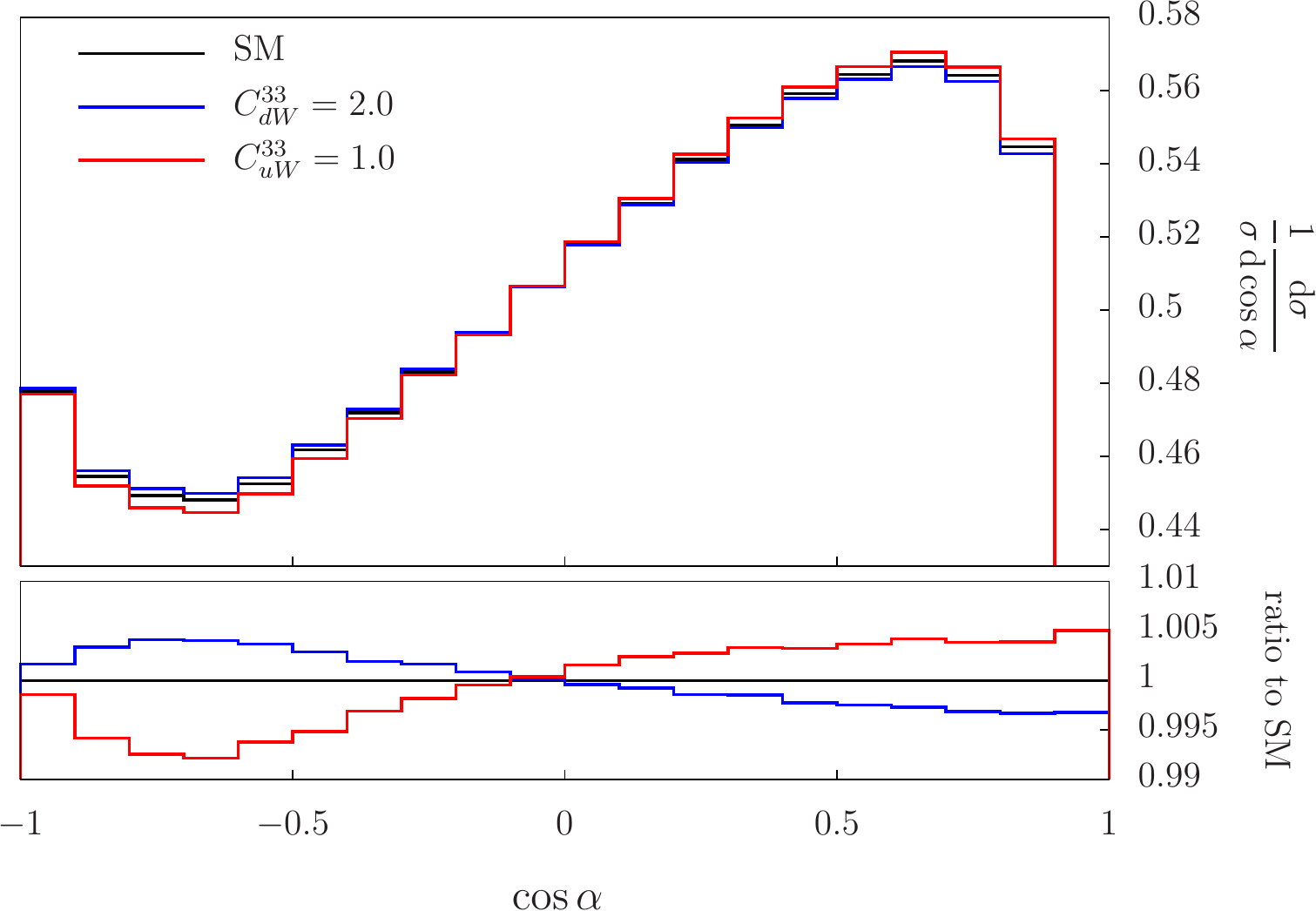}
 \caption{Angular distributions in $pp \to t \bar{t} \to b \bar{b} \, \ell \nu \, jj  $ at 13~TeV used to construct the asymmetries $A_{\theta_{\ell}^*}$ and $A_{\alpha}$.
 \label{fig:AngObs} }
\end{figure*}

In this section we expand and complement the constraints obtained from $t\bar{t}+\gamma$ and $t\bar{t}+Z$ 
in the previous section.
We make use of large $t\bar{t}$ cross section at the LHC to define asymmetries of angular distributions from 
the top quark decay products in order to constrain the remaining Wilson coefficient $C^{33}_{dW}$, and to
over-constrain $C^{33}_{uW}$.
As before, we use the lepton+jet final state of the $t\bar{t}$ system.
Similar ideas have been presented in Ref.~\cite{AguilarSaavedra:2010nx}, which puts more emphasis on 
probing the complex phases of the anomalous couplings.
Here, we consider the two angles, defined by
\begin{align}
\cos \theta_{\ell}^* = \frac{ \vec{\hat{p}}_\ell \cdot \vec{\tilde{p}}_W } {  |\vec{\hat{p}}_\ell|\, |\vec{\tilde{p}}_W| },
\quad\quad
\cos \alpha = \frac{ \vec{p}_t \cdot \vec{\tilde{p}}_W } {  |\vec{p}_t| \, |\vec{\tilde{p}}_W| },
\end{align}
where $\vec{\hat{p}}_\ell$ is the lepton momentum in the corresponding $W$ rest frame, $\vec{\tilde{p}}_W$ is the $W$ momentum in the corresponding top quark rest frame,
and $\vec{p}_t$ is the top quark momentum in the laboratory frame. 
Their kinematic distributions are shown in Fig.~\ref{fig:AngObs} for the SM and two anomalous coupling choices.
From these distributions we construct asymmetries
\begin{equation}
   \label{eq:Adef}
   A_\phi ({c_0}) = \frac{ \sigma(\cos\phi<c_0) - \sigma(\cos\phi>c_0) }{\sigma(\cos\phi<c_0)+\sigma(\cos\phi>c_0)},
\end{equation}
and use $A_{\theta_{\ell}^*}({ -0.1 })$ and $A_{\alpha}({0.0})$ 
to maximize their value in our analysis. 
We evaluated the SM asymmetries at NLO QCD and find perturbative shifts of only $\mathcal{O}(+0.5\%)$.
In our coupling analysis below we will assume a slightly inflated uncertainty of $\pm 4\%$.
On the experimental side, we expect a similar precision because of the large statistical sample at 13~TeV and 
the fact that existing measurements at 8~TeV already reach $10\%$ precision for spin asymmetries~\cite{Khachatryan:2016xws,Aad:2014mfk}. 

\begin{figure*}[ht]
 \begin{center}
 \includegraphics[width=0.32\textwidth]{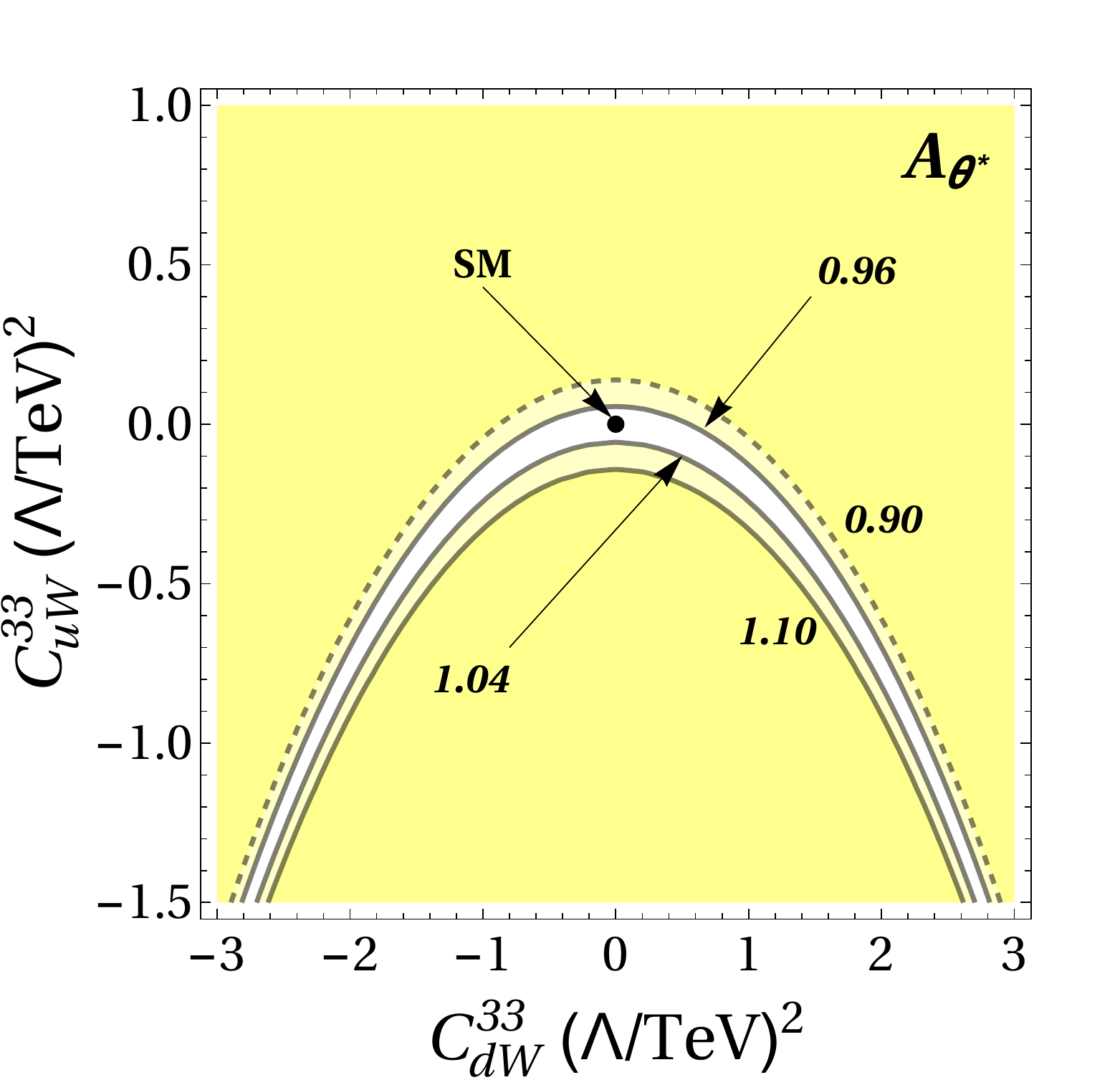} \hfill
 \includegraphics[width=0.32\textwidth]{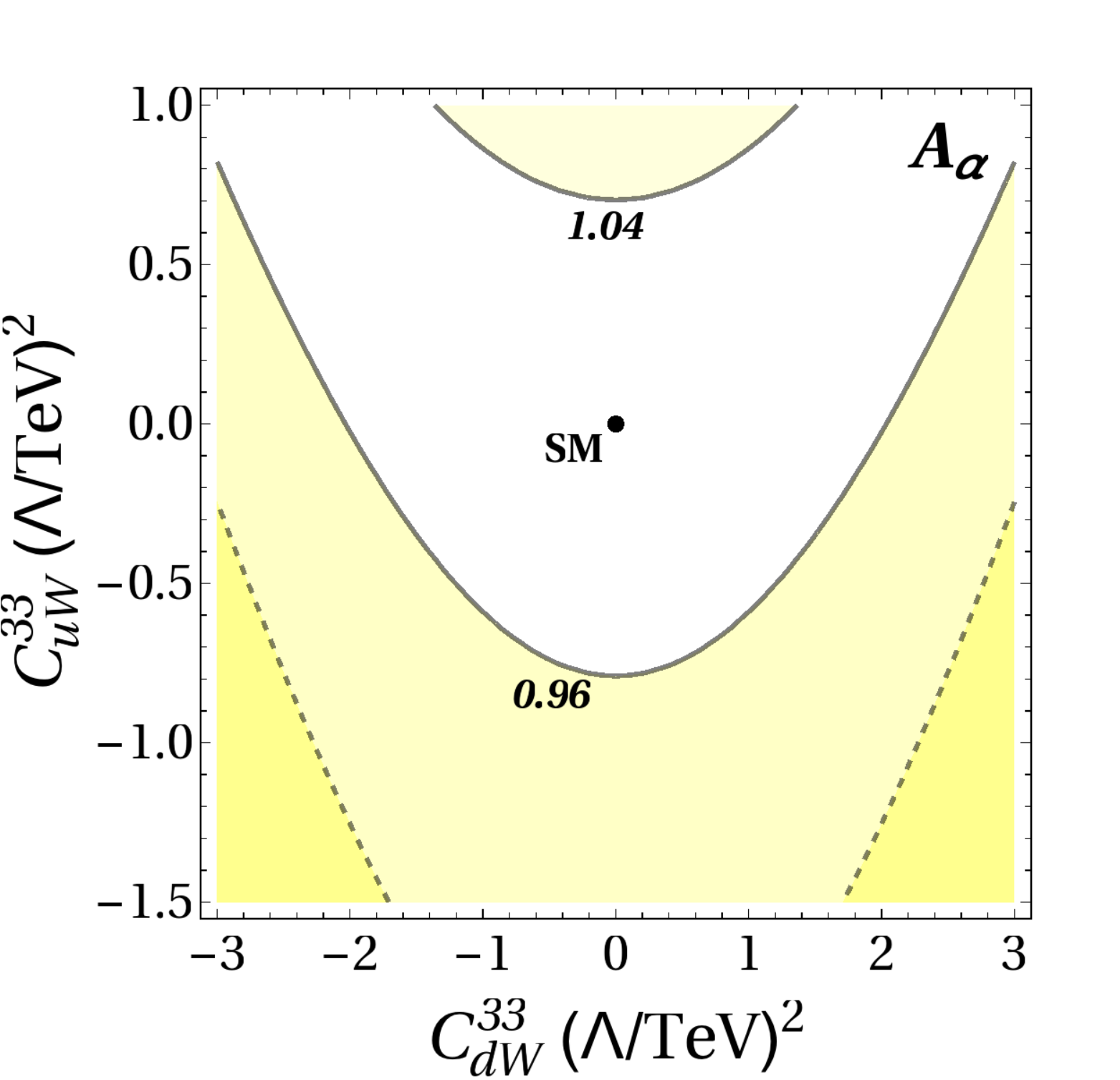} \hfill
 \includegraphics[width=0.32\textwidth]{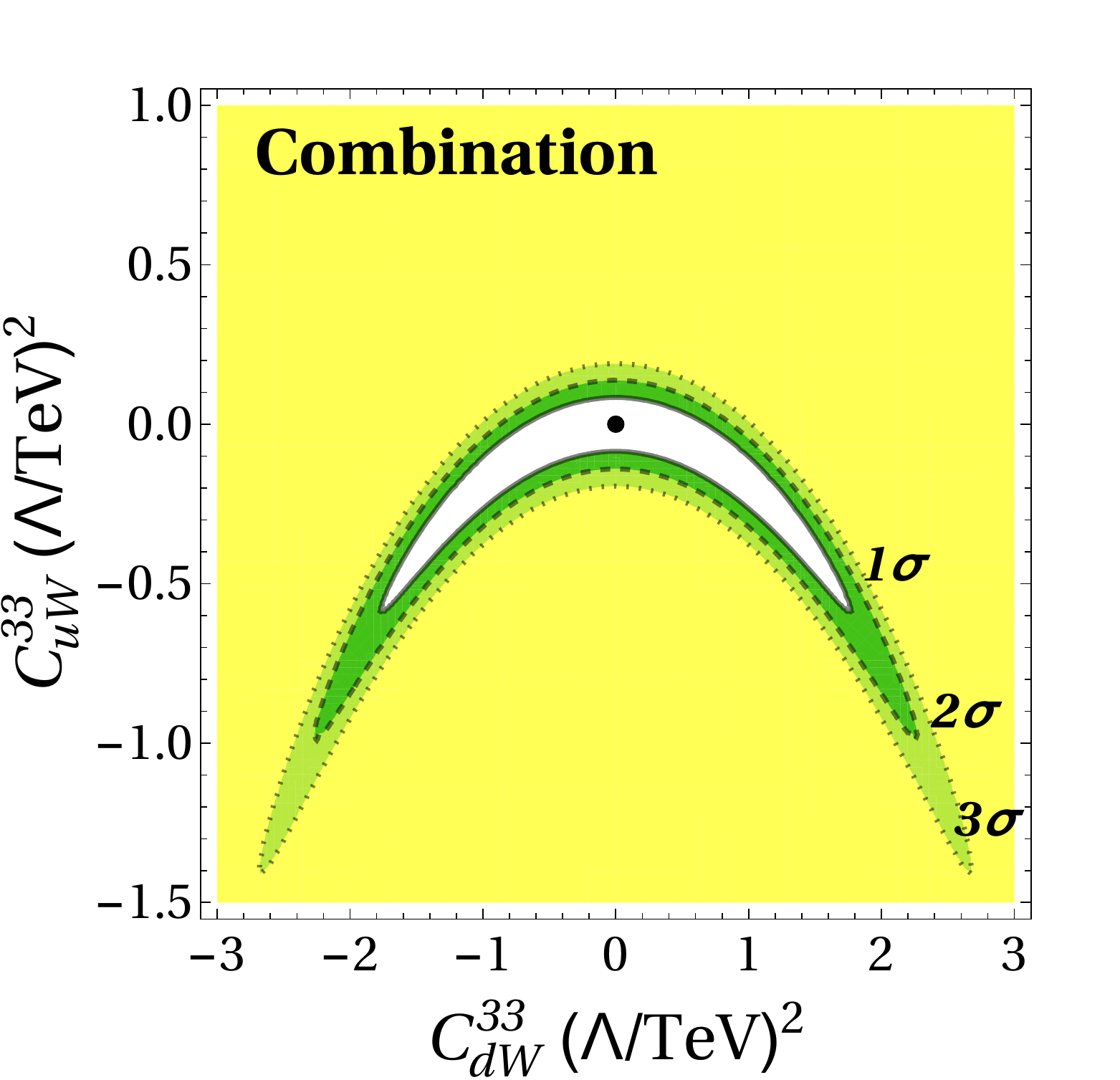} 
 \caption{Angular asymmetries $A_{\theta_{\ell}^*}(-0.1)$~(left) and $A_{\alpha}(0.0)$~(middle) as a function of the two Wilson coefficients in $t \bar{t}$ production relative to the their SM values. 
  Right: $\chi^2$ combination of the two asymmetries assuming an uncertainty of $\pm 4$\%.  }
 \label{fig:AngAsymm}
 \end{center}
\end{figure*}

To probe the sensitivity of the two asymmetries to the dipole couplings we vary $C^{33}_{dW}$ and $C^{33}_{uW}$ between $[-4,4]$ in steps of 0.5.
Hence, we perform 324 computations at leading order and sub-sequentially fit the results to an analytic parameterization that 
is shown in Fig.~\ref{fig:AngAsymm}.
Interestingly, the functional dependence of the asymmetries on the dipole operator couplings is opposite: 
$A_{\theta_{\ell}^*}$ has a concave shape (Fig.~\ref{fig:AngAsymm}, left) as a function of $C_{uW}^{33}$ and $C_{dW}^{33}$, 
whereas $A_{\alpha}$ (middle) is convexly shaped.
This feature allows us to combine the two constraints, shown on the right of Fig.~\ref{fig:AngAsymm},
in order to break the (white) invariance bands of the two separate observables.
As a result, both anomalous operators are clearly bounded in the combination plot. 
Moreover, the constraints on $C_{uW}^{33}$ can be further utilized in conjunction with the 
bounds obtained from cross section ratios (Fig.~\ref{fig:ratioconstraints}, right)
as they exclude almost the entire region of positive $C_{uW}^{33}$ values.
Turning the argument around, two independent measurements of 1) cross section ratios and 2) angular asymmetries 
can also be used to over-constrain the Wilson coefficient $C_{uW}^{33}$.

\section{Summary} \label{sec:summary}

In this paper we investigated the prospects of constraining 
electroweak dipole moments of the top quark at the 13 TeV LHC.
The SM radiatively generates these couplings through electroweak loop corrections, 
which turn out to be too small to be observed at the LHC.
This opens up the possibility to search for sizable deviations from zero as a way to probe new physics in the top quark sector.

We considered all anomalous dipole interactions between the top quark
and the electroweak gauge bosons in $\bar{t} b W$, $t\bar{t}\gamma$ and $t\bar{t}Z$ interactions,
and we showed that the processes $pp \to t\bar{t}$ and $pp \to t\bar{t}+\gamma/Z$ are
ideal probes for studying them. 
While these couplings enter simultaneously in various places,
electroweak gauge symmetries relate them and lead to only a small number of relevant operators. 
In order to constrain and disentangle them, we propose the study of
cross section ratios to make use of {\it orthogonal} sensitivity to anomalous operators entering $t\bar{t}\gamma$ and $t\bar{t}Z$.
We carefully investigated the ratios at NLO QCD and verify a strong reduction 
of uncertainties related to parton distribution functions and higher-order corrections. 
Experimental systematics such as luminosity or jet energy scale uncertainties are expected to drop out as well,
yielding true precision observables.

The final results of our study confirm this picture as we find that 13 TeV data can place firm bounds on the contributing operators.
Marginalizing over other operators, we find sensitivity of $C_{uW}^{33}=[-1.2,+1.4] \, (\Lambda/\mathrm{TeV})^2$ 
and $C_{uB\phi}^{33}=[-1.9,+1.2] \, (\Lambda/\mathrm{TeV})^2$ at the 95\% CL from combining the cross section ratios,
assuming that the theoretical accuracy of 3\,\% is matched by the experimental one.

We corroborate our results by studying angular asymmetries in $t\bar{t}$ production
to bound the last operator in our analysis and find the sensitivity 
$C_{dW}^{33}=[-2.0,+2.0]  \, (\Lambda/\mathrm{TeV})^2$.
Since the angular asymmetries are also sensitive to $C_{uW}^{33}$,
they can yield the independent constraint $C_{uW}^{33}=[-0.8,+0.1] \, (\Lambda/\mathrm{TeV})^2$ which 
can be further used to boost our results from cross section ratios.
Altogether, our proposal to construct four precision observables allows to 
pin down all of the three anomalous electroweak dipole operators
without a blind direction.

We note that for a 100\,TeV $pp$ collider the $t\bar{t}Z$ and $t\bar{t}\gamma$ cross sections are about a factor of 30 larger than at the 13\,TeV LHC. 
Thus, the statistical error will be sub-dominant only after a few tens fb$^{-1}$ of integrated luminosity. 
In order to fully exploit the potential of a 1\,ab$^{-1}$ data set, the theoretical predictions should be improved by 
one order of magnitude.  
Theoretical control of cross section ratios at the per-mille level may be possible once predictions at next-to-next-to-leading order QCD 
are available for $t\bar{t}+\gamma/Z$.
This does not seem unrealistic on the relevant time scale and would boost the constraints by more than 100\,\%.

Future work on improving our results could be a more precise understanding of uncertainties of the cross section ratios.
For example, it would be desirable to have the complete NLO predictions for the $t\bar{t}+\gamma$ process with anomalous couplings.
A fully realistic analysis also needs to consider backgrounds which we neglected in this work.
The incorporation of single-top quark processes into our analysis or more differential observables 
will certainly further strengthen sensitivity to new physics. 
\\

\acknowledgments{
We thank Michelangelo Mangano, Gilad Perez, Raoul R\"ontsch for comments on the manuscript and Jesse Thaler for helpful discussions.
M.S. is grateful for the hospitality at MIT-CTP where this project was initiated. 
Y.S. is supported by the U.S. Department of Energy (DOE) under cooperative research agreement DE-SC-00012567.
}

\bibliography{ref}

\begin{thebibliography}{77}
\expandafter\ifx\csname natexlab\endcsname\relax\def\natexlab#1{#1}\fi
\expandafter\ifx\csname bibnamefont\endcsname\relax
  \def\bibnamefont#1{#1}\fi
\expandafter\ifx\csname bibfnamefont\endcsname\relax
  \def\bibfnamefont#1{#1}\fi
\expandafter\ifx\csname citenamefont\endcsname\relax
  \def\citenamefont#1{#1}\fi
\expandafter\ifx\csname url\endcsname\relax
  \def\url#1{\texttt{#1}}\fi
\expandafter\ifx\csname urlprefix\endcsname\relax\def\urlprefix{URL }\fi
\providecommand{\bibinfo}[2]{#2}
\providecommand{\eprint}[2][]{\url{#2}}

\bibitem[{\citenamefont{Wicke}(2011)}]{Wicke:2010cg}
\bibinfo{author}{\bibfnamefont{D.}~\bibnamefont{Wicke}}, \bibinfo{journal}{Eur.
  Phys. J.} \textbf{\bibinfo{volume}{C71}}, \bibinfo{pages}{1627}
  (\bibinfo{year}{2011}), \eprint{1005.2460}.

\bibitem[{\citenamefont{Deliot and Glenzinski}(2012)}]{Deliot:2010ey}
\bibinfo{author}{\bibfnamefont{F.}~\bibnamefont{Deliot}} \bibnamefont{and}
  \bibinfo{author}{\bibfnamefont{D.~A.} \bibnamefont{Glenzinski}},
  \bibinfo{journal}{Rev. Mod. Phys.} \textbf{\bibinfo{volume}{84}},
  \bibinfo{pages}{211} (\bibinfo{year}{2012}), \eprint{1010.1202}.

\bibitem[{\citenamefont{Schilling}(2012)}]{Schilling:2012dx}
\bibinfo{author}{\bibfnamefont{F.-P.} \bibnamefont{Schilling}},
  \bibinfo{journal}{Int. J. Mod. Phys.} \textbf{\bibinfo{volume}{A27}},
  \bibinfo{pages}{1230016} (\bibinfo{year}{2012}), \eprint{1206.4484}.

\bibitem[{\citenamefont{del Duca and Laenen}(2015)}]{delDuca:2015gca}
\bibinfo{author}{\bibfnamefont{V.}~\bibnamefont{del Duca}} \bibnamefont{and}
  \bibinfo{author}{\bibfnamefont{E.}~\bibnamefont{Laenen}},
  \bibinfo{journal}{Int. J. Mod. Phys.} \textbf{\bibinfo{volume}{A30}},
  \bibinfo{pages}{1530063} (\bibinfo{year}{2015}), \eprint{1510.06690}.

\bibitem[{\citenamefont{Khachatryan et~al.}(2014{\natexlab{a}})}]{CMS:2014wma}
\bibinfo{author}{\bibfnamefont{V.}~\bibnamefont{Khachatryan}}
  \bibnamefont{et~al.} (\bibinfo{collaboration}{CMS}),
  \bibinfo{journal}{CMS-PAS-TOP-13-011}  (\bibinfo{year}{2014}{\natexlab{a}}).

\bibitem[{\citenamefont{Khachatryan
  et~al.}(2014{\natexlab{b}})}]{Khachatryan:2014qaa}
\bibinfo{author}{\bibfnamefont{V.}~\bibnamefont{Khachatryan}}
  \bibnamefont{et~al.} (\bibinfo{collaboration}{CMS}), \bibinfo{journal}{JHEP}
  \textbf{\bibinfo{volume}{09}}, \bibinfo{pages}{087}
  (\bibinfo{year}{2014}{\natexlab{b}}), \bibinfo{note}{[Erratum:
  JHEP10,106(2014)]}, \eprint{1408.1682}.

\bibitem[{\citenamefont{Aad et~al.}(2015{\natexlab{a}})}]{Aad:2015iha}
\bibinfo{author}{\bibfnamefont{G.}~\bibnamefont{Aad}} \bibnamefont{et~al.}
  (\bibinfo{collaboration}{ATLAS}), \bibinfo{journal}{Phys. Lett.}
  \textbf{\bibinfo{volume}{B749}}, \bibinfo{pages}{519}
  (\bibinfo{year}{2015}{\natexlab{a}}), \eprint{1506.05988}.

\bibitem[{\citenamefont{Aad et~al.}(2015{\natexlab{b}})}]{Aad:2015gra}
\bibinfo{author}{\bibfnamefont{G.}~\bibnamefont{Aad}} \bibnamefont{et~al.}
  (\bibinfo{collaboration}{ATLAS}), \bibinfo{journal}{Eur. Phys. J.}
  \textbf{\bibinfo{volume}{C75}}, \bibinfo{pages}{349}
  (\bibinfo{year}{2015}{\natexlab{b}}), \eprint{1503.05066}.

\bibitem[{\citenamefont{Aad et~al.}(2015{\natexlab{c}})}]{Aad:2014lma}
\bibinfo{author}{\bibfnamefont{G.}~\bibnamefont{Aad}} \bibnamefont{et~al.}
  (\bibinfo{collaboration}{ATLAS}), \bibinfo{journal}{Phys. Lett.}
  \textbf{\bibinfo{volume}{B740}}, \bibinfo{pages}{222}
  (\bibinfo{year}{2015}{\natexlab{c}}), \eprint{1409.3122}.

\bibitem[{\citenamefont{Aad et~al.}(2015{\natexlab{d}})}]{Aad:2015uwa}
\bibinfo{author}{\bibfnamefont{G.}~\bibnamefont{Aad}} \bibnamefont{et~al.}
  (\bibinfo{collaboration}{ATLAS}), \bibinfo{journal}{Phys. Rev.}
  \textbf{\bibinfo{volume}{D91}}, \bibinfo{pages}{072007}
  (\bibinfo{year}{2015}{\natexlab{d}}), \eprint{1502.00586}.

\bibitem[{\citenamefont{Khachatryan
  et~al.}(2015{\natexlab{a}})}]{Khachatryan:2015ila}
\bibinfo{author}{\bibfnamefont{V.}~\bibnamefont{Khachatryan}}
  \bibnamefont{et~al.} (\bibinfo{collaboration}{CMS}), \bibinfo{journal}{Eur.
  Phys. J.} \textbf{\bibinfo{volume}{C75}}, \bibinfo{pages}{251}
  (\bibinfo{year}{2015}{\natexlab{a}}), \eprint{1502.02485}.

\bibitem[{\citenamefont{Khachatryan
  et~al.}(2016{\natexlab{a}})}]{Khachatryan:2015sha}
\bibinfo{author}{\bibfnamefont{V.}~\bibnamefont{Khachatryan}}
  \bibnamefont{et~al.} (\bibinfo{collaboration}{CMS}), \bibinfo{journal}{JHEP}
  \textbf{\bibinfo{volume}{01}}, \bibinfo{pages}{096}
  (\bibinfo{year}{2016}{\natexlab{a}}), \eprint{1510.01131}.

\bibitem[{\citenamefont{Aad et~al.}(2015{\natexlab{e}})}]{Aad:2015eua}
\bibinfo{author}{\bibfnamefont{G.}~\bibnamefont{Aad}} \bibnamefont{et~al.}
  (\bibinfo{collaboration}{ATLAS}), \bibinfo{journal}{JHEP}
  \textbf{\bibinfo{volume}{11}}, \bibinfo{pages}{172}
  (\bibinfo{year}{2015}{\natexlab{e}}), \eprint{1509.05276}.

\bibitem[{\citenamefont{Aaltonen et~al.}(2012)}]{Aaltonen:2012rz}
\bibinfo{author}{\bibfnamefont{T.}~\bibnamefont{Aaltonen}} \bibnamefont{et~al.}
  (\bibinfo{collaboration}{CDF, D0}), \bibinfo{journal}{Phys. Rev.}
  \textbf{\bibinfo{volume}{D85}}, \bibinfo{pages}{071106}
  (\bibinfo{year}{2012}), \eprint{1202.5272}.

\bibitem[{\citenamefont{Khachatryan
  et~al.}(2015{\natexlab{b}})}]{Khachatryan:2014vma}
\bibinfo{author}{\bibfnamefont{V.}~\bibnamefont{Khachatryan}}
  \bibnamefont{et~al.} (\bibinfo{collaboration}{CMS}), \bibinfo{journal}{JHEP}
  \textbf{\bibinfo{volume}{01}}, \bibinfo{pages}{053}
  (\bibinfo{year}{2015}{\natexlab{b}}), \eprint{1410.1154}.

\bibitem[{\citenamefont{Aad et~al.}(2016)}]{Aad:2015bfa}
\bibinfo{author}{\bibfnamefont{G.}~\bibnamefont{Aad}} \bibnamefont{et~al.}
  (\bibinfo{collaboration}{ATLAS}), \bibinfo{journal}{Phys. Rev.}
  \textbf{\bibinfo{volume}{D93}}, \bibinfo{pages}{012002}
  (\bibinfo{year}{2016}), \eprint{1510.07478}.

\bibitem[{\citenamefont{Aad et~al.}(2015{\natexlab{f}})}]{Aad:2015yem}
\bibinfo{author}{\bibfnamefont{G.}~\bibnamefont{Aad}} \bibnamefont{et~al.}
  (\bibinfo{collaboration}{ATLAS}) (\bibinfo{year}{2015}{\natexlab{f}}),
  \eprint{1510.03764}.

\bibitem[{\citenamefont{Aad et~al.}(2011)}]{ATLAS:2011dha}
\bibinfo{author}{\bibfnamefont{G.}~\bibnamefont{Aad}} \bibnamefont{et~al.}
  (\bibinfo{collaboration}{ATLAS}), \bibinfo{journal}{ATLAS-CONF-2011-141}
  (\bibinfo{year}{2011}).

\bibitem[{\citenamefont{Khachatryan et~al.}(2012)}]{CMS:2012oua}
\bibinfo{author}{\bibfnamefont{V.}~\bibnamefont{Khachatryan}}
  \bibnamefont{et~al.} (\bibinfo{collaboration}{CMS}),
  \bibinfo{journal}{CMS-PAS-TOP-11-031}  (\bibinfo{year}{2012}).

\bibitem[{\citenamefont{Aaltonen et~al.}(2013)}]{Aaltonen:2013sgl}
\bibinfo{author}{\bibfnamefont{T.}~\bibnamefont{Aaltonen}} \bibnamefont{et~al.}
  (\bibinfo{collaboration}{CDF}), \bibinfo{journal}{Phys. Rev.}
  \textbf{\bibinfo{volume}{D88}}, \bibinfo{pages}{032003}
  (\bibinfo{year}{2013}), \eprint{1304.4141}.

\bibitem[{\citenamefont{Abazov et~al.}(2014)}]{Abazov:2014lha}
\bibinfo{author}{\bibfnamefont{V.~M.} \bibnamefont{Abazov}}
  \bibnamefont{et~al.} (\bibinfo{collaboration}{D0}), \bibinfo{journal}{Phys.
  Rev.} \textbf{\bibinfo{volume}{D90}}, \bibinfo{pages}{051101}
  (\bibinfo{year}{2014}), \bibinfo{note}{[Erratum: Phys.
  Rev.D90,no.7,079904(2014)]}, \eprint{1407.4837}.

\bibitem[{\citenamefont{Grzadkowski and Misiak}(2008)}]{Grzadkowski:2008mf}
\bibinfo{author}{\bibfnamefont{B.}~\bibnamefont{Grzadkowski}} \bibnamefont{and}
  \bibinfo{author}{\bibfnamefont{M.}~\bibnamefont{Misiak}},
  \bibinfo{journal}{Phys. Rev.} \textbf{\bibinfo{volume}{D78}},
  \bibinfo{pages}{077501} (\bibinfo{year}{2008}), \bibinfo{note}{[Erratum:
  Phys. Rev.D84,059903(2011)]}, \eprint{0802.1413}.

\bibitem[{\citenamefont{Kamenik et~al.}(2012)\citenamefont{Kamenik, Papucci,
  and Weiler}}]{Kamenik:2011dk}
\bibinfo{author}{\bibfnamefont{J.~F.} \bibnamefont{Kamenik}},
  \bibinfo{author}{\bibfnamefont{M.}~\bibnamefont{Papucci}}, \bibnamefont{and}
  \bibinfo{author}{\bibfnamefont{A.}~\bibnamefont{Weiler}},
  \bibinfo{journal}{Phys. Rev.} \textbf{\bibinfo{volume}{D85}},
  \bibinfo{pages}{071501} (\bibinfo{year}{2012}), \bibinfo{note}{[Erratum:
  Phys. Rev.D88,no.3,039903(2013)]}, \eprint{1107.3143}.

\bibitem[{\citenamefont{Brod et~al.}(2015)\citenamefont{Brod, Greljo, Stamou,
  and Uttayarat}}]{Brod:2014hsa}
\bibinfo{author}{\bibfnamefont{J.}~\bibnamefont{Brod}},
  \bibinfo{author}{\bibfnamefont{A.}~\bibnamefont{Greljo}},
  \bibinfo{author}{\bibfnamefont{E.}~\bibnamefont{Stamou}}, \bibnamefont{and}
  \bibinfo{author}{\bibfnamefont{P.}~\bibnamefont{Uttayarat}},
  \bibinfo{journal}{JHEP} \textbf{\bibinfo{volume}{02}}, \bibinfo{pages}{141}
  (\bibinfo{year}{2015}), \eprint{1408.0792}.

\bibitem[{\citenamefont{Schael et~al.}(2006)}]{ALEPH:2005ab}
\bibinfo{author}{\bibfnamefont{S.}~\bibnamefont{Schael}} \bibnamefont{et~al.}
  (\bibinfo{collaboration}{SLD Electroweak Group, DELPHI, ALEPH, SLD, SLD Heavy
  Flavour Group, OPAL, LEP Electroweak Working Group, L3}),
  \bibinfo{journal}{Phys. Rept.} \textbf{\bibinfo{volume}{427}},
  \bibinfo{pages}{257} (\bibinfo{year}{2006}), \eprint{hep-ex/0509008}.

\bibitem[{\citenamefont{Abdallah et~al.}(2009)}]{Abdallah:2008ab}
\bibinfo{author}{\bibfnamefont{J.}~\bibnamefont{Abdallah}} \bibnamefont{et~al.}
  (\bibinfo{collaboration}{DELPHI}), \bibinfo{journal}{Eur. Phys. J.}
  \textbf{\bibinfo{volume}{C60}}, \bibinfo{pages}{1} (\bibinfo{year}{2009}),
  \eprint{0901.4461}.

\bibitem[{\citenamefont{Larios et~al.}(1999)\citenamefont{Larios, Perez, and
  Yuan}}]{Larios:1999au}
\bibinfo{author}{\bibfnamefont{F.}~\bibnamefont{Larios}},
  \bibinfo{author}{\bibfnamefont{M.~A.} \bibnamefont{Perez}}, \bibnamefont{and}
  \bibinfo{author}{\bibfnamefont{C.~P.} \bibnamefont{Yuan}},
  \bibinfo{journal}{Phys. Lett.} \textbf{\bibinfo{volume}{B457}},
  \bibinfo{pages}{334} (\bibinfo{year}{1999}), \eprint{hep-ph/9903394}.

\bibitem[{\citenamefont{de~Blas et~al.}(2015)\citenamefont{de~Blas, Chala, and
  Santiago}}]{deBlas:2015aea}
\bibinfo{author}{\bibfnamefont{J.}~\bibnamefont{de~Blas}},
  \bibinfo{author}{\bibfnamefont{M.}~\bibnamefont{Chala}}, \bibnamefont{and}
  \bibinfo{author}{\bibfnamefont{J.}~\bibnamefont{Santiago}},
  \bibinfo{journal}{JHEP} \textbf{\bibinfo{volume}{09}}, \bibinfo{pages}{189}
  (\bibinfo{year}{2015}), \eprint{1507.00757}.

\bibitem[{\citenamefont{Baur et~al.}(2005)\citenamefont{Baur, Juste, Orr, and
  Rainwater}}]{Baur:2004uw}
\bibinfo{author}{\bibfnamefont{U.}~\bibnamefont{Baur}},
  \bibinfo{author}{\bibfnamefont{A.}~\bibnamefont{Juste}},
  \bibinfo{author}{\bibfnamefont{L.~H.} \bibnamefont{Orr}}, \bibnamefont{and}
  \bibinfo{author}{\bibfnamefont{D.}~\bibnamefont{Rainwater}},
  \bibinfo{journal}{Phys. Rev.} \textbf{\bibinfo{volume}{D71}},
  \bibinfo{pages}{054013} (\bibinfo{year}{2005}), \eprint{hep-ph/0412021}.

\bibitem[{\citenamefont{Berger et~al.}(2009)\citenamefont{Berger, Cao, and
  Low}}]{Berger:2009hi}
\bibinfo{author}{\bibfnamefont{E.~L.} \bibnamefont{Berger}},
  \bibinfo{author}{\bibfnamefont{Q.-H.} \bibnamefont{Cao}}, \bibnamefont{and}
  \bibinfo{author}{\bibfnamefont{I.}~\bibnamefont{Low}},
  \bibinfo{journal}{Phys. Rev.} \textbf{\bibinfo{volume}{D80}},
  \bibinfo{pages}{074020} (\bibinfo{year}{2009}), \eprint{0907.2191}.

\bibitem[{\citenamefont{Aguilar-Saavedra and
  Bernabeu}(2010)}]{AguilarSaavedra:2010nx}
\bibinfo{author}{\bibfnamefont{J.~A.} \bibnamefont{Aguilar-Saavedra}}
  \bibnamefont{and} \bibinfo{author}{\bibfnamefont{J.}~\bibnamefont{Bernabeu}},
  \bibinfo{journal}{Nucl. Phys.} \textbf{\bibinfo{volume}{B840}},
  \bibinfo{pages}{349} (\bibinfo{year}{2010}), \eprint{1005.5382}.

\bibitem[{\citenamefont{Aguilar-Saavedra
  et~al.}(2011)\citenamefont{Aguilar-Saavedra, Castro, and
  Onofre}}]{AguilarSaavedra:2011ct}
\bibinfo{author}{\bibfnamefont{J.~A.} \bibnamefont{Aguilar-Saavedra}},
  \bibinfo{author}{\bibfnamefont{N.~F.} \bibnamefont{Castro}},
  \bibnamefont{and} \bibinfo{author}{\bibfnamefont{A.}~\bibnamefont{Onofre}},
  \bibinfo{journal}{Phys. Rev.} \textbf{\bibinfo{volume}{D83}},
  \bibinfo{pages}{117301} (\bibinfo{year}{2011}), \eprint{1105.0117}.

\bibitem[{\citenamefont{Rindani and Sharma}(2011)}]{Rindani:2011pk}
\bibinfo{author}{\bibfnamefont{S.~D.} \bibnamefont{Rindani}} \bibnamefont{and}
  \bibinfo{author}{\bibfnamefont{P.}~\bibnamefont{Sharma}},
  \bibinfo{journal}{JHEP} \textbf{\bibinfo{volume}{11}}, \bibinfo{pages}{082}
  (\bibinfo{year}{2011}), \eprint{1107.2597}.

\bibitem[{\citenamefont{Bach and Ohl}(2012)}]{Bach:2012fb}
\bibinfo{author}{\bibfnamefont{F.}~\bibnamefont{Bach}} \bibnamefont{and}
  \bibinfo{author}{\bibfnamefont{T.}~\bibnamefont{Ohl}},
  \bibinfo{journal}{Phys. Rev.} \textbf{\bibinfo{volume}{D86}},
  \bibinfo{pages}{114026} (\bibinfo{year}{2012}), \eprint{1209.4564}.

\bibitem[{\citenamefont{Zhang et~al.}(2012)\citenamefont{Zhang, Greiner, and
  Willenbrock}}]{Zhang:2012cd}
\bibinfo{author}{\bibfnamefont{C.}~\bibnamefont{Zhang}},
  \bibinfo{author}{\bibfnamefont{N.}~\bibnamefont{Greiner}}, \bibnamefont{and}
  \bibinfo{author}{\bibfnamefont{S.}~\bibnamefont{Willenbrock}},
  \bibinfo{journal}{Phys. Rev.} \textbf{\bibinfo{volume}{D86}},
  \bibinfo{pages}{014024} (\bibinfo{year}{2012}), \eprint{1201.6670}.

\bibitem[{\citenamefont{Aguilar-Saavedra and dos
  Santos}(2014)}]{Aguilar-Saavedra:2014eqa}
\bibinfo{author}{\bibfnamefont{J.~A.} \bibnamefont{Aguilar-Saavedra}}
  \bibnamefont{and} \bibinfo{author}{\bibfnamefont{S.~A.} \bibnamefont{dos
  Santos}}, \bibinfo{journal}{Phys. Rev.} \textbf{\bibinfo{volume}{D89}},
  \bibinfo{pages}{114009} (\bibinfo{year}{2014}), \eprint{1404.1585}.

\bibitem[{\citenamefont{Zhang}(2014)}]{Zhang:2014rja}
\bibinfo{author}{\bibfnamefont{C.}~\bibnamefont{Zhang}},
  \bibinfo{journal}{Phys. Rev.} \textbf{\bibinfo{volume}{D90}},
  \bibinfo{pages}{014008} (\bibinfo{year}{2014}), \eprint{1404.1264}.

\bibitem[{\citenamefont{Durieux et~al.}(2015)\citenamefont{Durieux, Maltoni,
  and Zhang}}]{Durieux:2014xla}
\bibinfo{author}{\bibfnamefont{G.}~\bibnamefont{Durieux}},
  \bibinfo{author}{\bibfnamefont{F.}~\bibnamefont{Maltoni}}, \bibnamefont{and}
  \bibinfo{author}{\bibfnamefont{C.}~\bibnamefont{Zhang}},
  \bibinfo{journal}{Phys. Rev.} \textbf{\bibinfo{volume}{D91}},
  \bibinfo{pages}{074017} (\bibinfo{year}{2015}), \eprint{1412.7166}.

\bibitem[{\citenamefont{Prasath~V et~al.}(2015)\citenamefont{Prasath~V,
  Godbole, and Rindani}}]{Prasath:2014mfa}
\bibinfo{author}{\bibfnamefont{A.}~\bibnamefont{Prasath~V}},
  \bibinfo{author}{\bibfnamefont{R.~M.} \bibnamefont{Godbole}},
  \bibnamefont{and} \bibinfo{author}{\bibfnamefont{S.~D.}
  \bibnamefont{Rindani}}, \bibinfo{journal}{Eur. Phys. J.}
  \textbf{\bibinfo{volume}{C75}}, \bibinfo{pages}{402} (\bibinfo{year}{2015}),
  \eprint{1405.1264}.

\bibitem[{\citenamefont{Aguilar-Saavedra
  et~al.}(2014)\citenamefont{Aguilar-Saavedra, Álvarez, Juste, and
  Rubbo}}]{Aguilar-Saavedra:2014vta}
\bibinfo{author}{\bibfnamefont{J.~A.} \bibnamefont{Aguilar-Saavedra}},
  \bibinfo{author}{\bibfnamefont{E.}~\bibnamefont{Álvarez}},
  \bibinfo{author}{\bibfnamefont{A.}~\bibnamefont{Juste}}, \bibnamefont{and}
  \bibinfo{author}{\bibfnamefont{F.}~\bibnamefont{Rubbo}},
  \bibinfo{journal}{JHEP} \textbf{\bibinfo{volume}{04}}, \bibinfo{pages}{188}
  (\bibinfo{year}{2014}), \eprint{1402.3598}.

\bibitem[{\citenamefont{Bernardo et~al.}(2014)\citenamefont{Bernardo, Castro,
  Fiolhais, Gonçalves, Guerra, Oliveira, and Onofre}}]{Bernardo:2014vha}
\bibinfo{author}{\bibfnamefont{C.}~\bibnamefont{Bernardo}},
  \bibinfo{author}{\bibfnamefont{N.~F.} \bibnamefont{Castro}},
  \bibinfo{author}{\bibfnamefont{M.~C.~N.} \bibnamefont{Fiolhais}},
  \bibinfo{author}{\bibfnamefont{H.}~\bibnamefont{Gonçalves}},
  \bibinfo{author}{\bibfnamefont{A.~G.~C.} \bibnamefont{Guerra}},
  \bibinfo{author}{\bibfnamefont{M.}~\bibnamefont{Oliveira}}, \bibnamefont{and}
  \bibinfo{author}{\bibfnamefont{A.}~\bibnamefont{Onofre}},
  \bibinfo{journal}{Phys. Rev.} \textbf{\bibinfo{volume}{D90}},
  \bibinfo{pages}{113007} (\bibinfo{year}{2014}), \eprint{1408.7063}.

\bibitem[{\citenamefont{Rontsch and Schulze}(2014)}]{Rontsch:2014cca}
\bibinfo{author}{\bibfnamefont{R.}~\bibnamefont{Rontsch}} \bibnamefont{and}
  \bibinfo{author}{\bibfnamefont{M.}~\bibnamefont{Schulze}},
  \bibinfo{journal}{JHEP} \textbf{\bibinfo{volume}{07}}, \bibinfo{pages}{091}
  (\bibinfo{year}{2014}), \bibinfo{note}{[Erratum: JHEP09,132(2015)]},
  \eprint{1404.1005}.

\bibitem[{\citenamefont{Tonero and Rosenfeld}(2014)}]{Tonero:2014jea}
\bibinfo{author}{\bibfnamefont{A.}~\bibnamefont{Tonero}} \bibnamefont{and}
  \bibinfo{author}{\bibfnamefont{R.}~\bibnamefont{Rosenfeld}},
  \bibinfo{journal}{Phys. Rev.} \textbf{\bibinfo{volume}{D90}},
  \bibinfo{pages}{017701} (\bibinfo{year}{2014}), \eprint{1404.2581}.

\bibitem[{\citenamefont{Rontsch and Schulze}(2015)}]{Rontsch:2015una}
\bibinfo{author}{\bibfnamefont{R.}~\bibnamefont{Rontsch}} \bibnamefont{and}
  \bibinfo{author}{\bibfnamefont{M.}~\bibnamefont{Schulze}},
  \bibinfo{journal}{JHEP} \textbf{\bibinfo{volume}{08}}, \bibinfo{pages}{044}
  (\bibinfo{year}{2015}), \eprint{1501.05939}.

\bibitem[{\citenamefont{Buckley et~al.}(2015)\citenamefont{Buckley, Englert,
  Ferrando, Miller, Moore, Russell, and White}}]{Buckley:2015lku}
\bibinfo{author}{\bibfnamefont{A.}~\bibnamefont{Buckley}},
  \bibinfo{author}{\bibfnamefont{C.}~\bibnamefont{Englert}},
  \bibinfo{author}{\bibfnamefont{J.}~\bibnamefont{Ferrando}},
  \bibinfo{author}{\bibfnamefont{D.~J.} \bibnamefont{Miller}},
  \bibinfo{author}{\bibfnamefont{L.}~\bibnamefont{Moore}},
  \bibinfo{author}{\bibfnamefont{M.}~\bibnamefont{Russell}}, \bibnamefont{and}
  \bibinfo{author}{\bibfnamefont{C.~D.} \bibnamefont{White}}
  (\bibinfo{year}{2015}), \eprint{1512.03360}.

\bibitem[{\citenamefont{Zhang}(2016)}]{Zhang:2016omx}
\bibinfo{author}{\bibfnamefont{C.}~\bibnamefont{Zhang}} (\bibinfo{year}{2016}),
  \eprint{1601.06163}.

\bibitem[{\citenamefont{Bylund et~al.}(2016)\citenamefont{Bylund, Maltoni,
  Tsinikos, Vryonidou, and Zhang}}]{Bylund:2016phk}
\bibinfo{author}{\bibfnamefont{O.~B.} \bibnamefont{Bylund}},
  \bibinfo{author}{\bibfnamefont{F.}~\bibnamefont{Maltoni}},
  \bibinfo{author}{\bibfnamefont{I.}~\bibnamefont{Tsinikos}},
  \bibinfo{author}{\bibfnamefont{E.}~\bibnamefont{Vryonidou}},
  \bibnamefont{and} \bibinfo{author}{\bibfnamefont{C.}~\bibnamefont{Zhang}}
  (\bibinfo{year}{2016}), \eprint{1601.08193}.

\bibitem[{\citenamefont{Cao et~al.}(2015)\citenamefont{Cao, Yan, Yu, and
  Zhang}}]{Cao:2015doa}
\bibinfo{author}{\bibfnamefont{Q.-H.} \bibnamefont{Cao}},
  \bibinfo{author}{\bibfnamefont{B.}~\bibnamefont{Yan}},
  \bibinfo{author}{\bibfnamefont{J.-H.} \bibnamefont{Yu}}, \bibnamefont{and}
  \bibinfo{author}{\bibfnamefont{C.}~\bibnamefont{Zhang}}
  (\bibinfo{year}{2015}), \eprint{1504.03785}.

\bibitem[{\citenamefont{Cirigliano et~al.}(2016)\citenamefont{Cirigliano,
  Dekens, de~Vries, and Mereghetti}}]{Cirigliano:2016njn}
\bibinfo{author}{\bibfnamefont{V.}~\bibnamefont{Cirigliano}},
  \bibinfo{author}{\bibfnamefont{W.}~\bibnamefont{Dekens}},
  \bibinfo{author}{\bibfnamefont{J.}~\bibnamefont{de~Vries}}, \bibnamefont{and}
  \bibinfo{author}{\bibfnamefont{E.}~\bibnamefont{Mereghetti}}
  (\bibinfo{year}{2016}), \eprint{1603.03049}.

\bibitem[{\citenamefont{Aguilar-Saavedra}(2009)}]{AguilarSaavedra:2008zc}
\bibinfo{author}{\bibfnamefont{J.~A.} \bibnamefont{Aguilar-Saavedra}},
  \bibinfo{journal}{Nucl. Phys.} \textbf{\bibinfo{volume}{B812}},
  \bibinfo{pages}{181} (\bibinfo{year}{2009}), \eprint{0811.3842}.

\bibitem[{\citenamefont{Bernabeu et~al.}(1996)\citenamefont{Bernabeu, Comelli,
  Lavoura, and Silva}}]{Bernabeu:1995gs}
\bibinfo{author}{\bibfnamefont{J.}~\bibnamefont{Bernabeu}},
  \bibinfo{author}{\bibfnamefont{D.}~\bibnamefont{Comelli}},
  \bibinfo{author}{\bibfnamefont{L.}~\bibnamefont{Lavoura}}, \bibnamefont{and}
  \bibinfo{author}{\bibfnamefont{J.~P.} \bibnamefont{Silva}},
  \bibinfo{journal}{Phys. Rev.} \textbf{\bibinfo{volume}{D53}},
  \bibinfo{pages}{5222} (\bibinfo{year}{1996}), \eprint{hep-ph/9509416}.

\bibitem[{\citenamefont{Czarnecki and Krause}(1997)}]{Czarnecki:1996rx}
\bibinfo{author}{\bibfnamefont{A.}~\bibnamefont{Czarnecki}} \bibnamefont{and}
  \bibinfo{author}{\bibfnamefont{B.}~\bibnamefont{Krause}},
  \bibinfo{journal}{Acta Phys. Polon.} \textbf{\bibinfo{volume}{B28}},
  \bibinfo{pages}{829} (\bibinfo{year}{1997}), \eprint{hep-ph/9611299}.

\bibitem[{\citenamefont{Hollik et~al.}(1999)\citenamefont{Hollik, Illana,
  Rigolin, Schappacher, and Stockinger}}]{Hollik:1998vz}
\bibinfo{author}{\bibfnamefont{W.}~\bibnamefont{Hollik}},
  \bibinfo{author}{\bibfnamefont{J.~I.} \bibnamefont{Illana}},
  \bibinfo{author}{\bibfnamefont{S.}~\bibnamefont{Rigolin}},
  \bibinfo{author}{\bibfnamefont{C.}~\bibnamefont{Schappacher}},
  \bibnamefont{and}
  \bibinfo{author}{\bibfnamefont{D.}~\bibnamefont{Stockinger}},
  \bibinfo{journal}{Nucl. Phys.} \textbf{\bibinfo{volume}{B551}},
  \bibinfo{pages}{3} (\bibinfo{year}{1999}), \bibinfo{note}{[Erratum: Nucl.
  Phys.B557,407(1999)]}, \eprint{hep-ph/9812298}.

\bibitem[{\citenamefont{Agashe et~al.}(2007)\citenamefont{Agashe, Perez, and
  Soni}}]{Agashe:2006wa}
\bibinfo{author}{\bibfnamefont{K.}~\bibnamefont{Agashe}},
  \bibinfo{author}{\bibfnamefont{G.}~\bibnamefont{Perez}}, \bibnamefont{and}
  \bibinfo{author}{\bibfnamefont{A.}~\bibnamefont{Soni}},
  \bibinfo{journal}{Phys. Rev.} \textbf{\bibinfo{volume}{D75}},
  \bibinfo{pages}{015002} (\bibinfo{year}{2007}), \eprint{hep-ph/0606293}.

\bibitem[{\citenamefont{Kagan et~al.}(2009)\citenamefont{Kagan, Perez,
  Volansky, and Zupan}}]{Kagan:2009bn}
\bibinfo{author}{\bibfnamefont{A.~L.} \bibnamefont{Kagan}},
  \bibinfo{author}{\bibfnamefont{G.}~\bibnamefont{Perez}},
  \bibinfo{author}{\bibfnamefont{T.}~\bibnamefont{Volansky}}, \bibnamefont{and}
  \bibinfo{author}{\bibfnamefont{J.}~\bibnamefont{Zupan}},
  \bibinfo{journal}{Phys. Rev.} \textbf{\bibinfo{volume}{D80}},
  \bibinfo{pages}{076002} (\bibinfo{year}{2009}), \eprint{0903.1794}.

\bibitem[{\citenamefont{Ibrahim and Nath}(2010)}]{Ibrahim:2010hv}
\bibinfo{author}{\bibfnamefont{T.}~\bibnamefont{Ibrahim}} \bibnamefont{and}
  \bibinfo{author}{\bibfnamefont{P.}~\bibnamefont{Nath}},
  \bibinfo{journal}{Phys. Rev.} \textbf{\bibinfo{volume}{D82}},
  \bibinfo{pages}{055001} (\bibinfo{year}{2010}), \eprint{1007.0432}.

\bibitem[{\citenamefont{Ibrahim and Nath}(2011)}]{Ibrahim:2011im}
\bibinfo{author}{\bibfnamefont{T.}~\bibnamefont{Ibrahim}} \bibnamefont{and}
  \bibinfo{author}{\bibfnamefont{P.}~\bibnamefont{Nath}},
  \bibinfo{journal}{Phys. Rev.} \textbf{\bibinfo{volume}{D84}},
  \bibinfo{pages}{015003} (\bibinfo{year}{2011}), \eprint{1104.3851}.

\bibitem[{\citenamefont{Grojean et~al.}(2013)\citenamefont{Grojean,
  Matsedonskyi, and Panico}}]{Grojean:2013qca}
\bibinfo{author}{\bibfnamefont{C.}~\bibnamefont{Grojean}},
  \bibinfo{author}{\bibfnamefont{O.}~\bibnamefont{Matsedonskyi}},
  \bibnamefont{and} \bibinfo{author}{\bibfnamefont{G.}~\bibnamefont{Panico}},
  \bibinfo{journal}{JHEP} \textbf{\bibinfo{volume}{10}}, \bibinfo{pages}{160}
  (\bibinfo{year}{2013}), \eprint{1306.4655}.

\bibitem[{\citenamefont{Richard}(2013)}]{Richard:2013pwa}
\bibinfo{author}{\bibfnamefont{F.}~\bibnamefont{Richard}}
  (\bibinfo{year}{2013}), \eprint{1304.3594}.

\bibitem[{\citenamefont{Melnikov et~al.}(2011)\citenamefont{Melnikov, Schulze,
  and Scharf}}]{Melnikov:2011ta}
\bibinfo{author}{\bibfnamefont{K.}~\bibnamefont{Melnikov}},
  \bibinfo{author}{\bibfnamefont{M.}~\bibnamefont{Schulze}}, \bibnamefont{and}
  \bibinfo{author}{\bibfnamefont{A.}~\bibnamefont{Scharf}},
  \bibinfo{journal}{Phys. Rev.} \textbf{\bibinfo{volume}{D83}},
  \bibinfo{pages}{074013} (\bibinfo{year}{2011}), \eprint{1102.1967}.

\bibitem[{\citenamefont{Fox et~al.}(2008)\citenamefont{Fox, Ligeti, Papucci,
  Perez, and Schwartz}}]{Fox:2007in}
\bibinfo{author}{\bibfnamefont{P.~J.} \bibnamefont{Fox}},
  \bibinfo{author}{\bibfnamefont{Z.}~\bibnamefont{Ligeti}},
  \bibinfo{author}{\bibfnamefont{M.}~\bibnamefont{Papucci}},
  \bibinfo{author}{\bibfnamefont{G.}~\bibnamefont{Perez}}, \bibnamefont{and}
  \bibinfo{author}{\bibfnamefont{M.~D.} \bibnamefont{Schwartz}},
  \bibinfo{journal}{Phys. Rev.} \textbf{\bibinfo{volume}{D78}},
  \bibinfo{pages}{054008} (\bibinfo{year}{2008}), \eprint{0704.1482}.

\bibitem[{\citenamefont{Drobnak et~al.}(2012)\citenamefont{Drobnak, Fajfer, and
  Kamenik}}]{Drobnak:2011aa}
\bibinfo{author}{\bibfnamefont{J.}~\bibnamefont{Drobnak}},
  \bibinfo{author}{\bibfnamefont{S.}~\bibnamefont{Fajfer}}, \bibnamefont{and}
  \bibinfo{author}{\bibfnamefont{J.~F.} \bibnamefont{Kamenik}},
  \bibinfo{journal}{Nucl. Phys.} \textbf{\bibinfo{volume}{B855}},
  \bibinfo{pages}{82} (\bibinfo{year}{2012}), \eprint{1109.2357}.

\bibitem[{\citenamefont{Aguilar-Saavedra
  et~al.}(2015)\citenamefont{Aguilar-Saavedra, Fuks, and
  Mangano}}]{Aguilar-Saavedra:2014iga}
\bibinfo{author}{\bibfnamefont{J.~A.} \bibnamefont{Aguilar-Saavedra}},
  \bibinfo{author}{\bibfnamefont{B.}~\bibnamefont{Fuks}}, \bibnamefont{and}
  \bibinfo{author}{\bibfnamefont{M.~L.} \bibnamefont{Mangano}},
  \bibinfo{journal}{Phys. Rev.} \textbf{\bibinfo{volume}{D91}},
  \bibinfo{pages}{094021} (\bibinfo{year}{2015}), \eprint{1412.6654}.

\bibitem[{\citenamefont{Bernreuther et~al.}(1992)\citenamefont{Bernreuther,
  Schroder, and Pham}}]{Bernreuther:1992dz}
\bibinfo{author}{\bibfnamefont{W.}~\bibnamefont{Bernreuther}},
  \bibinfo{author}{\bibfnamefont{T.}~\bibnamefont{Schroder}}, \bibnamefont{and}
  \bibinfo{author}{\bibfnamefont{T.~N.} \bibnamefont{Pham}},
  \bibinfo{journal}{Phys. Lett.} \textbf{\bibinfo{volume}{B279}},
  \bibinfo{pages}{389} (\bibinfo{year}{1992}).

\bibitem[{\citenamefont{Hernandez-Sanchez
  et~al.}(2007)\citenamefont{Hernandez-Sanchez, Procopio, Tavares-Velasco, and
  Toscano}}]{HernandezSanchez:2006sw}
\bibinfo{author}{\bibfnamefont{J.}~\bibnamefont{Hernandez-Sanchez}},
  \bibinfo{author}{\bibfnamefont{F.}~\bibnamefont{Procopio}},
  \bibinfo{author}{\bibfnamefont{G.}~\bibnamefont{Tavares-Velasco}},
  \bibnamefont{and} \bibinfo{author}{\bibfnamefont{J.~J.}
  \bibnamefont{Toscano}}, \bibinfo{journal}{Phys. Rev.}
  \textbf{\bibinfo{volume}{D75}}, \bibinfo{pages}{073017}
  (\bibinfo{year}{2007}), \eprint{hep-ph/0611379}.

\bibitem[{\citenamefont{Dekens et~al.}(2014)\citenamefont{Dekens, de~Vries,
  Bsaisou, Bernreuther, Hanhart, Mei\ss{}ner, Nogga, and
  Wirzba}}]{Dekens:2014jka}
\bibinfo{author}{\bibfnamefont{W.}~\bibnamefont{Dekens}},
  \bibinfo{author}{\bibfnamefont{J.}~\bibnamefont{de~Vries}},
  \bibinfo{author}{\bibfnamefont{J.}~\bibnamefont{Bsaisou}},
  \bibinfo{author}{\bibfnamefont{W.}~\bibnamefont{Bernreuther}},
  \bibinfo{author}{\bibfnamefont{C.}~\bibnamefont{Hanhart}},
  \bibinfo{author}{\bibfnamefont{U.-G.} \bibnamefont{Mei\ss{}ner}},
  \bibinfo{author}{\bibfnamefont{A.}~\bibnamefont{Nogga}}, \bibnamefont{and}
  \bibinfo{author}{\bibfnamefont{A.}~\bibnamefont{Wirzba}},
  \bibinfo{journal}{JHEP} \textbf{\bibinfo{volume}{07}}, \bibinfo{pages}{069}
  (\bibinfo{year}{2014}), \eprint{1404.6082}.

\bibitem[{\citenamefont{Schulze and R\"ontsch}(2014)}]{RefTOPAZ}
\bibinfo{author}{\bibfnamefont{M.}~\bibnamefont{Schulze}} \bibnamefont{and}
  \bibinfo{author}{\bibfnamefont{R.}~\bibnamefont{R\"ontsch}},
  \emph{\bibinfo{title}{Topaz}},
  \bibinfo{howpublished}{\url{https://github.com/TOPAZdevelop/TOPAZ}}
  (\bibinfo{year}{2014}).

\bibitem[{\citenamefont{Melnikov and Schulze}(2009)}]{Melnikov:2009dn}
\bibinfo{author}{\bibfnamefont{K.}~\bibnamefont{Melnikov}} \bibnamefont{and}
  \bibinfo{author}{\bibfnamefont{M.}~\bibnamefont{Schulze}},
  \bibinfo{journal}{JHEP} \textbf{\bibinfo{volume}{08}}, \bibinfo{pages}{049}
  (\bibinfo{year}{2009}), \eprint{0907.3090}.

\bibitem[{\citenamefont{Cacciari et~al.}(2008)\citenamefont{Cacciari, Salam,
  and Soyez}}]{Cacciari:2008gp}
\bibinfo{author}{\bibfnamefont{M.}~\bibnamefont{Cacciari}},
  \bibinfo{author}{\bibfnamefont{G.~P.} \bibnamefont{Salam}}, \bibnamefont{and}
  \bibinfo{author}{\bibfnamefont{G.}~\bibnamefont{Soyez}},
  \bibinfo{journal}{JHEP} \textbf{\bibinfo{volume}{04}}, \bibinfo{pages}{063}
  (\bibinfo{year}{2008}), \eprint{0802.1189}.

\bibitem[{\citenamefont{Ball et~al.}(2015)}]{Ball:2014uwa}
\bibinfo{author}{\bibfnamefont{R.~D.} \bibnamefont{Ball}} \bibnamefont{et~al.}
  (\bibinfo{collaboration}{NNPDF}), \bibinfo{journal}{JHEP}
  \textbf{\bibinfo{volume}{04}}, \bibinfo{pages}{040} (\bibinfo{year}{2015}),
  \eprint{1410.8849}.

\bibitem[{\citenamefont{Mangano et~al.}(2016)\citenamefont{Mangano, Plehn,
  Reimitz, Schell, and Shao}}]{Plehn:2015cta}
\bibinfo{author}{\bibfnamefont{M.~L.} \bibnamefont{Mangano}},
  \bibinfo{author}{\bibfnamefont{T.}~\bibnamefont{Plehn}},
  \bibinfo{author}{\bibfnamefont{P.}~\bibnamefont{Reimitz}},
  \bibinfo{author}{\bibfnamefont{T.}~\bibnamefont{Schell}}, \bibnamefont{and}
  \bibinfo{author}{\bibfnamefont{H.-S.} \bibnamefont{Shao}},
  \bibinfo{journal}{J. Phys.} \textbf{\bibinfo{volume}{G43}},
  \bibinfo{pages}{035001} (\bibinfo{year}{2016}), \eprint{1507.08169}.

\bibitem[{\citenamefont{Pumplin et~al.}(2002)\citenamefont{Pumplin, Stump,
  Huston, Lai, Nadolsky, and Tung}}]{Pumplin:2002vw}
\bibinfo{author}{\bibfnamefont{J.}~\bibnamefont{Pumplin}},
  \bibinfo{author}{\bibfnamefont{D.~R.} \bibnamefont{Stump}},
  \bibinfo{author}{\bibfnamefont{J.}~\bibnamefont{Huston}},
  \bibinfo{author}{\bibfnamefont{H.~L.} \bibnamefont{Lai}},
  \bibinfo{author}{\bibfnamefont{P.~M.} \bibnamefont{Nadolsky}},
  \bibnamefont{and} \bibinfo{author}{\bibfnamefont{W.~K.} \bibnamefont{Tung}},
  \bibinfo{journal}{JHEP} \textbf{\bibinfo{volume}{07}}, \bibinfo{pages}{012}
  (\bibinfo{year}{2002}), \eprint{hep-ph/0201195}.

\bibitem[{\citenamefont{Martin et~al.}(2009)\citenamefont{Martin, Stirling,
  Thorne, and Watt}}]{Martin:2009iq}
\bibinfo{author}{\bibfnamefont{A.~D.} \bibnamefont{Martin}},
  \bibinfo{author}{\bibfnamefont{W.~J.} \bibnamefont{Stirling}},
  \bibinfo{author}{\bibfnamefont{R.~S.} \bibnamefont{Thorne}},
  \bibnamefont{and} \bibinfo{author}{\bibfnamefont{G.}~\bibnamefont{Watt}},
  \bibinfo{journal}{Eur. Phys. J.} \textbf{\bibinfo{volume}{C63}},
  \bibinfo{pages}{189} (\bibinfo{year}{2009}), \eprint{0901.0002}.

\bibitem[{\citenamefont{Kuhn et~al.}(2015)\citenamefont{Kuhn, Scharf, and
  Uwer}}]{Kuhn:2013zoa}
\bibinfo{author}{\bibfnamefont{J.~H.} \bibnamefont{Kuhn}},
  \bibinfo{author}{\bibfnamefont{A.}~\bibnamefont{Scharf}}, \bibnamefont{and}
  \bibinfo{author}{\bibfnamefont{P.}~\bibnamefont{Uwer}},
  \bibinfo{journal}{Phys. Rev.} \textbf{\bibinfo{volume}{D91}},
  \bibinfo{pages}{014020} (\bibinfo{year}{2015}), \eprint{1305.5773}.

\bibitem[{\citenamefont{Frixione et~al.}(2015)\citenamefont{Frixione, Hirschi,
  Pagani, Shao, and Zaro}}]{Frixione:2015zaa}
\bibinfo{author}{\bibfnamefont{S.}~\bibnamefont{Frixione}},
  \bibinfo{author}{\bibfnamefont{V.}~\bibnamefont{Hirschi}},
  \bibinfo{author}{\bibfnamefont{D.}~\bibnamefont{Pagani}},
  \bibinfo{author}{\bibfnamefont{H.~S.} \bibnamefont{Shao}}, \bibnamefont{and}
  \bibinfo{author}{\bibfnamefont{M.}~\bibnamefont{Zaro}},
  \bibinfo{journal}{JHEP} \textbf{\bibinfo{volume}{06}}, \bibinfo{pages}{184}
  (\bibinfo{year}{2015}), \eprint{1504.03446}.

\bibitem[{\citenamefont{Khachatryan
  et~al.}(2016{\natexlab{b}})}]{Khachatryan:2016xws}
\bibinfo{author}{\bibfnamefont{V.}~\bibnamefont{Khachatryan}}
  \bibnamefont{et~al.} (\bibinfo{collaboration}{CMS})
  (\bibinfo{year}{2016}{\natexlab{b}}), \eprint{1601.01107}.

\bibitem[{\citenamefont{Aad et~al.}(2015{\natexlab{g}})}]{Aad:2014mfk}
\bibinfo{author}{\bibfnamefont{G.}~\bibnamefont{Aad}} \bibnamefont{et~al.}
  (\bibinfo{collaboration}{ATLAS}), \bibinfo{journal}{Phys. Rev. Lett.}
  \textbf{\bibinfo{volume}{114}}, \bibinfo{pages}{142001}
  (\bibinfo{year}{2015}{\natexlab{g}}), \eprint{1412.4742}.

\end{thebibliography}

\end{document}